\documentclass[twocolumn]{aastex631}
\usepackage{amsmath}
\usepackage{natbib}
\usepackage{xcolor}
\usepackage{comment}
\let\tablenum\relax
\usepackage{siunitx}

\newcommand{\kepler}{{\em Kepler}}
\newcommand{\tess}{{\em TESS}}
\newcommand{\gaia}{{\em Gaia}}
\newcommand{\gmag}{{\em G}}
\newcommand{\varindex}{{$varindex$}}

\begin{document}
\title{New Variable Hot Subdwarf Stars Identified from Anomalous \gaia\ Flux Errors,\\ Observed by \tess, and Classified via Fourier Diagnostics}


\author[0000-0002-8558-4353]{Brad~N.~Barlow}\affiliation{Department of Physics and Astronomy, High Point University, High Point, NC, USA}
\author[0000-0002-2764-7248]{Kyle~A.~Corcoran}\affiliation{Department of Astronomy, University of Virginia, Charlottesville, VA, USA}
\author[0000-0002-5280-6888]{Isabelle~M.~Parker}\affiliation{Department of Physics and Astronomy, High Point University, High Point, NC, USA}
\author[0000-0002-6540-1484]{Thomas~Kupfer}\affiliation{Department of Physics \& Astronomy, Texas Tech University, Lubbock, TX, USA}
\author[0000-0003-0963-0239]{P\'{e}ter~N\'{e}meth}\affiliation{Astronomical Institute of the Czech Academy of Sciences, 25165 Ondřejov, Czech Republic}\affiliation{Astroserver.org, 8533 Malomsok, Hungary}
\author[0000-0001-5941-2286]{J.~J.~Hermes}\affiliation{Department of Astronomy, Boston University, 725 Commonwealth Ave., Boston, MA 02215, USA}
\author[0000-0002-0009-409X]{Isaac~D.~Lopez}\affiliation{Department of Physics and Astronomy, High Point University, High Point, NC, USA}
\author[0000-0001-6027-9136]{Will~J.~Frondorf}\affiliation{Department of Physics and Astronomy, High Point University, High Point, NC, USA}
\author[0000-0003-2754-9418]{David~Vestal}\affiliation{Department of Physics and Astronomy, High Point University, High Point, NC, USA}
\author[0000-0002-9476-5471]{Jazzmyn~Holden}\affiliation{Department of Physics and Astronomy, High Point University, High Point, NC, USA}




\begin{abstract}
Hot subdwarf stars are mostly stripped red giants that can exhibit photometric variations due to stellar pulsations, eclipses, the reflection effect, ellipsoidal modulation, and Doppler beaming. Detailed studies of their light curves help constrain stellar parameters through asteroseismological analyses or binary light curve modeling and generally improve our capacity to draw a statistically meaningful picture of this enigmatic stage of stellar evolution.  
From an analysis of \gaia\ DR2 flux errors, we have identified around 1200 candidate hot subdwarfs with inflated flux errors for their magnitudes --- a strong indicator of photometric variability. As a pilot study, we obtained 2-min cadence \tess\ Cycle 2 observations of 187 candidate hot subdwarfs with anomalous \gaia\ flux errors. More than 90\% of our targets show significant photometric variations in their \tess\ light curves. Many of the new systems found are cataclysmic variables, but we report the discovery of several new variable hot subdwarfs, including HW Vir binaries, reflection effect systems, pulsating sdBV$_s$ stars, and ellipsoidally modulated systems. We determine atmospheric parameters for select systems using follow--up spectroscopy from the 3--m Shane telescope. Finally, we present a Fourier diagnostic plot for classifying binary light curves using the relative amplitudes and phases of their fundamental and harmonic signals in their periodograms.  This plot makes it possible to identify certain types of variables efficiently, without directly investigating their light curves, and may assist in the rapid classification of systems observed in large photometric surveys.\\
\end{abstract}

\section{Introduction} \label{sec:intro}

Hot subdwarf stars are hot, compact objects that are underluminous for their high temperatures. They can be broken into two spectroscopic classes: the sdB stars, which have temperatures from $20{,}000-40{,}000$\,K, and the sdO stars, which have temperatures greater than $40{,}000$\,K. Their spectra, dominated by H and He lines, show similarities with their main sequence O/B star counterparts but display much broader absorption profiles, indicative of their high surface gravities with log $g$ = $5.0 - 6.0$\,dex. Numerous evolutionary scenarios bring different types of stars near or through this part of the color-magnitude diagram, including extended horizontal branch (EHB) stars, pre-EHB stars, post-EHB stars, post-blue horizontal branch (BHB) stars, post-asymptotic giant branch (AGB) stars, and pre-helium white dwarfs (pre-He WD). Some cataclysmic variables (CVs) with significant accretion disk luminosities can even display similar colors and luminosities. The sdO stars, which are approximately one-third as prevalent as sdB stars, display a much broader array of properties and evolutonary histories. For a thorough review of hot subdwarf stars, see \citet{heb16}.

The majority of sdB stars are core He-burning EHB stars and the descendents of red giants that were stripped of their outer H envelopes. Binarity is the most widely accepted mechanism for their formation, with models invoking various common envelope and Roche lobe overflow interactions to explain their existence \citep{han02,han03}. The tightest sdB binaries have orbital periods as short as 30 min, with companions including brown dwarfs, red dwarfs, and even white dwarfs. Those with brown dwarf and red dwarf companions are identified photometrically from a strong reflection effect or eclipses. They likely formed through common envelope evolution, and studying their properties may shed light on the effects substellar companions have on stellar evolution, and on the mass--radius relation of hot Jupiters (e.g., \citealt{sch19}). Some of the shortest--period sdB+WD systems are candidate progenitors for type 1a supernovae (SN Ia) and gravitational wave verification sources for missions like LISA (e.g. \citealt{pel21}). Their light curves can show strong ellipsoidal modulations, Doppler beaming, eclipses, and, in the case of extremely high precision photometry, even white dwarf reflection and gravitational lensing. Finally, some hot subdwarfs, whether known to be in binaries or not, exhibit pulsations that permit deep probing with asteroseismology (e.g., \citealt{uzu21}). The rapidly--pulsating sdBV$_{\mathrm r}$\footnote{also known as V361 Hya, EC 14026, or sdBV$_{\mathrm p}$ stars; see \citet{kil10}} stars are pressure--mode (p--mode) pulsators with periods from 1-10 min and amplitudes typically $<$10 ppt. The cooler, slowly--pulsating sdBV$_{\mathrm s}$\footnote{also known as V1093 Her, Betsy, PG 1716, or sdBV$_{\mathrm g}$ stars; see \citet{kil10}} stars are gravity--mode (g--mode) pulsators with periods ranging from 1-2 hr and amplitudes usually $<$1 ppt. In the region where the hot edge of the sdBV$_{\mathrm s}$ instability strip overlaps the cool edge of the sdBV$_{\mathrm r}$ strip, several `hybrid' sdBV$_{\mathrm rs}$ stars have been observed with both p-- and g--mode oscillations \citep{bar05,sch06,lut09}. Careful monitoring of the frequency and phase modulations of hot subdwarf pulsations can lead to constraints on evolutionary rates or even the discovery of planetary-sized companions (e.g., \citealt{sch10, bar15}).

There is therefore significant value in expanding the numbers of known variable hot subdwarfs due to pulsations or binarity. Ground-based surveys for new variables using traditional telescopes can be slow and inefficient. Typically, only one target can be observed at a time, and significant time is spent observing stars that are not variable. Systems like the Zwicky Transient Facility (ZTF; \citealt{bel19}) and Evryscope \citep{law15} have proven to be invaluable for the discovery of new variables, with their large fields of view and high-cadence observations. But even these systems can suffer from weather-related problems, daytime gaps, and inadequate cadences. Space--based observatories like \kepler\ and the Transiting Exoplanet Survey Satellite (\tess) have greatly improved the situation, providing long, mostly uninterupted datasets of unprecented quality. \tess\ is particulary well--suited for time--domain studies of hot subdwarfs given the combination of its large field of view, rapid 2-min cadence mode (and 20-s cadence with Cycle 3 onward), and the intrinsic luminosity of hot subdwarfs which makes \tess\ suitable enough to deliver useful photometry. Due to limited bandwidth, however, 2-min cadence observations cannot be downloaded for {\em all} stars observed by \tess\ and must be targeted before observations.

In order to improve observing efficiency, we have utilized a technique to identify new variable hot subdwarf stars from \gaia\ DR2 data by selecting hot subdwarfs with anomalously large mean flux errors at a given mean magnitude. \tess\ Cycle 2 observations show this method effectively establishes variable candidates; more than 90\% of targets with anomalous flux errors have proven to be bona fide variables. Here we present the details of our target selection method (\S \ref{sec:selection}), a summary of the \tess\ Cycle 2 observations and analysis methods (\S \ref{sec:photometry}), a Fourier diagnostic tool that can aid in light curve classification (\S \ref{sec:fourier}), follow--up spectroscopy (\S \ref{sec:spectroscopy}), a breakdown of survey results (\S \ref{sec:results}), and a summary (\S \ref{sec:summary}).

\section{Candidate Variables from \\Anomalous Gaia Flux Errors} 
\label{sec:selection}

\begin{figure*}[t]
\centering
\includegraphics[width=2.1\columnwidth]{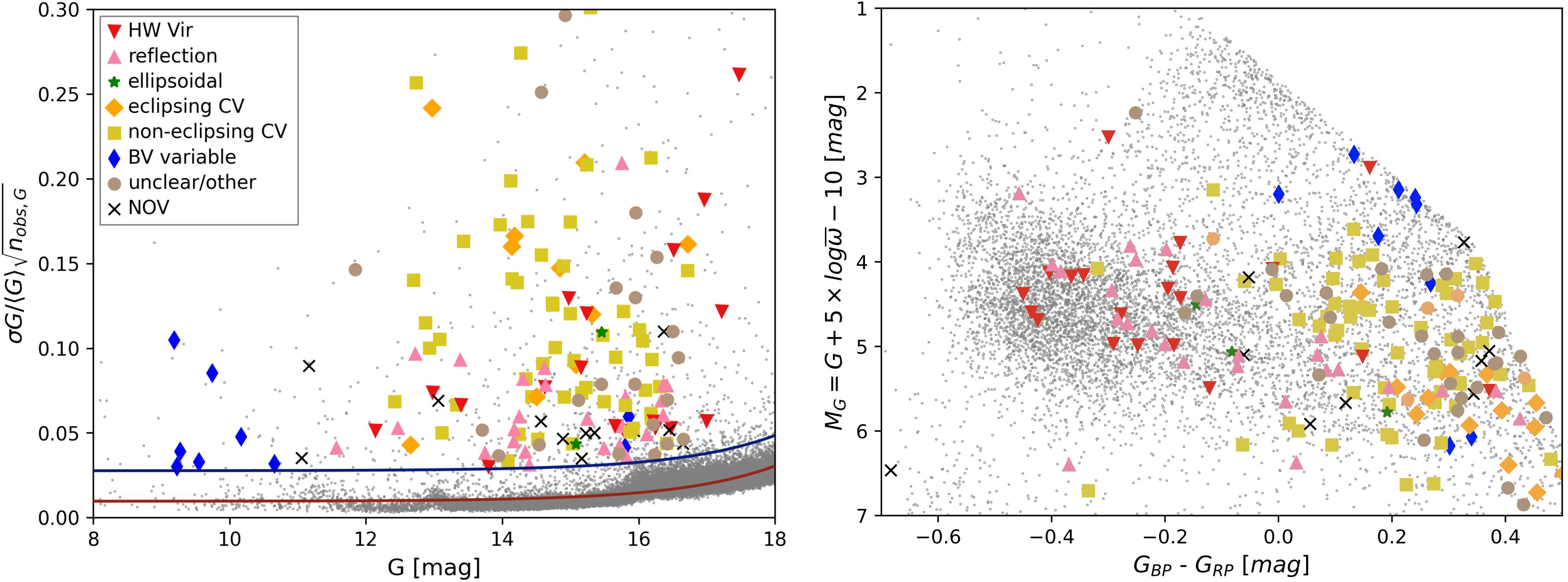} 

\caption{{\em Left panel:} Normalized \gaia\ \gmag\ flux errors plotted against \gmag\ magnitudes for all candidate hot subdwarf stars (gray points) from \citet{gei19}. The vast majority of systems fall along an exponential curve (red line) with normalized \gmag\ flux error increasing with decreasing flux. However, many show anomalously high flux errors for their magnitudes, an indication that they might be variables. We identify any systems falling more than 0.02 normalized flux error units above the exponential fit (blue line) as strong variable candidates. Systems with anomalouly high \gaia\ flux errors observed in our \tess\ Cycle 2 campaign are highlighted and color--coded according to their classification. Some of the highest \varindex\ objects extend beyond the top edge of the plot. {\em Right panel:} \gaia\ color-absolute magnitude diagram for all hot subdwarf candidates from \citet{gei19} with our observed systems highlighted using the same color--coding.}
\label{fig:varindex}
\end{figure*}

All of our targets are sourced from \citet{gei19}, who compiled a catalog of $39{,}800$ confirmed and candidate hot subdwarf stars from a variety of photometric and spectroscopic sources. More than 80\% of the stars were photometrically selected from their positions in the \gaia\ DR2 color--absolute magnitude diagram (hereafter, \gaia\ CMD), and from reduced proper motion cuts. With the exceptions of the Galactic plane and Magellanic clouds, they state the catalog is essentially complete out to $\sim$1.5\,kpc. Cuts were made to avoid significant contamination from white dwarfs (WDs) and low--mass main sequence stars, and they expect the majority of the candidates to be sdB and sdO hot subdwarfs, followed by late B-type BHB stars, hot post--AGB stars, and central stars of planetary nebulae (CSPN). They predict contamination by cooler stars to be around 10\%.

Variable stars can be identified effectively from the high-precision photometry published in \gaia\ DR2, without the need for a full light curve. Photometric uncertainties are empirically determined from the scatter of individual $G$ magnitude measurements (see \citealt{eva18} for full details).  Thus, variable stars should have higher--than--expected uncertainties compared to constant stars of comparable magnitude and number of observations. In order to put targets on an even playing field, we normalize the $G$ flux errors by taking the published values from \gaia\ DR2 ($\sigma _G$), dividing by the mean $G$ flux ($\langle G \rangle$), and multiplying by the square root of the number of $G$ flux observations ($n_{\mathrm{obs,G}}$). This \gaia\ variability metric is then described by:

\begin{equation}
V_G =  \frac{\sigma_G}{\langle G \rangle} \times \sqrt{n_{\mathrm{obs,G}}} 
\end{equation}
\noindent Further details regarding this method are presented by \citet{gui20}, who demonstrate its efficacy by applying it to a sample of $12{,}000$ white dwarfs.

In the left panel of Figure \ref{fig:varindex}, we plot for all hot subdwarfs in \citet{gei19} their \gaia\ variability metrics against their $G$ magnitudes (gray points).  The vast majority of our systems fall along a growing exponential curve, and their positions in the diagram are consistent with little--to--no variability\footnote{The ``bumpy'' features seen at G$\sim$13 and G$\sim$16 are related to changes in the window class of the observations, as explained in \citet{eva18}.}. More than a thousand systems, however, show anomalously high flux errors for their magnitudes, indicative of their photometric variability. We rank candidate variables based off their variability index (\varindex), which we define as their residuals after subtracting the best--fitting exponential curve, given by 

\begin{equation}
    varindex = V_G - (A \, e^{\beta G} + c)
\end{equation}

\noindent where $A$=$1.3466\times 10^{-8}$, $\beta=0.77634$, and $c=0.00962$. Systems with the highest \varindex\ values are most likely to be variable. We select as our strongest candidate variables all systems with \varindex\ $>$ 0.02, which corresponds roughly to the 3\% highest \varindex\ systems. Around 1200 hot subdwarf candidates from \citet{gei19} meet this criterion, and they are shown in Figure \ref{fig:varindex} and listed in Table 1. We note that the 22 months of photometry presented in DR2 was iteratively 5--sigma clipped, so it is possible the most variable or deeply eclipsing hot subdwarf systems might not exceed our \varindex\ treshhold. Additionally, many known sdBV stars show extremely small variations ($<$ 1 ppt) that would result in \varindex\ values below our detection threshold. For these reasons, we stress that our \gaia\ DR2 anomalous flux error method only establishes variability candidates --- it is incomplete at best. In a future work, we plan to explore and quantify the effects of light curve shape, orbital period, and other quanities on \varindex.

\begin{table*}
\label{tab:varindex}
\centering
{\begin{tabular}{llllllllll}
\hline
\gaia\ DR2 ID & Alias & RAJ2000 & DecJ2000 & $G$ & M$_{G}$ & G$_{BP}$-G$_{RP}$ & V$_{G}$ & \varindex \\
\hline
5822261072394880384 &  & 239.5537775 & -66.4043057 & 16.729 & 6.971 & 0.304 & 1.115 & 1.096\\
5205559736382852864 &  & 146.272336 & -74.46152364 & 17.870 & 6.291 & 0.384 & 1.071 & 1.043\\
2105585421693855744 & V*V363Lyr & 287.2149513 & 43.00868831 & 17.277 & 6.539 & 0.490 & 1.022 & 0.999\\
6182506024165282816 &  & 199.4152172 & -30.01218859 & 15.145 & 6.186 & 0.444 & 1.003 & 0.990\\
2116887915892603520 & V*V344Lyr & 281.1632287 & 43.37445875 & 16.298 & 6.164 & -0.063 & 1.002 & 0.985\\
5445756851859162752 &  & 155.5218803 & -35.63222379 & 15.467 & 5.533 & 0.254 & 0.903 & 0.890\\
3323819257020229888 &  & 96.06763184 & 5.611120955 & 13.822 & 4.547 & 0.387 & 0.870 & 0.860\\
6158701528704596096 &  & 191.3476054 & -35.16566666 & 17.008 & 5.904 & 0.248 & 0.816 & 0.796\\
4493803975199926144 &  & 262.5822517 & 12.04785217 & 16.500 & 6.426 & 0.512 & 0.807 & 0.789\\
795244943252669568 & V*RZLMi & 147.9538598 & 34.1233247 & 15.216 & 5.908 & 0.019 & 0.759 & 0.746\\
\ldots & \ldots & \ldots & \ldots & \ldots &\ldots & \ldots & \ldots & \ldots\\
\hline
\end{tabular}}
\caption{Known and candidate hot subdwarf stars from the \citet{gei19} catalog with the highest \varindex\ values, arranged from highest to lowest \varindex\ values. All systems included here have anomalously large \gaia\ flux errors indicating photometric variability. Full table is available online.}
\end{table*}

Several types of variable stars that are not bona fide hot subdwarfs fall within the same part of the \gaia\ CMD as the candidates selected by \citet{gei19}, and thus should be expected as contaminants in our sample. \citet{gai19variables} highlights the positions of nearly all known types of variable stars in the \gaia\ CMD in their Figures 3--7. At the highest luminosities and reddest colors, our sample could be contaminated by rapidly oscillating Am and Ap stars, slowly pulsating B stars (SPB), rotating B stars, $\gamma$ Doradus variables, $\delta$ Scuti, and even $\alpha ^2$ Canum Venaticorum stars. Cataclysmic variables likely pollute our sample at redder colors and lower luminosities. A study disentagling cataclysmic variables (CVs) in the \gaia\ CMD by \citet{abr20} shows we should expect to find numerous dwarf novae (DN), nova--like (NL) CVs, and intermediate polars (IP) in our color-selection region.  

\section{TESS Cycle 2 Photometry} \label{sec:photometry}
The \tess\ spacecraft \citep{ric15} concluded its Prime Mission on July 5, 2020, after conducting a two--year photometric survey of the southern and northern ecliptic hemispheres using four small telescopes covering a 24$^{\circ}$ $\times$ 96$^{\circ}$ strip of sky. Observations were divided into 26 observing sectors, each lasting 27 d.  While targets at low ecliptic latitudes were observed for at least 27 d continuously, those located at high ecliptic latitudes in the continuous viewing zone could be observed for much longer. Full frame images (FFI) were collected every 30 min, downloaded to Earth, and made publicly available. 2--min cadence observations available for approximately 20,000 stars in each sector and selected from the scientific community via calls for proposals. 

We obtained 2--min cadence \tess\ Cycle 2 observations for all bright ($G<17$\,mag), high--\varindex\ hot subdwarfs falling in Sectors 14--26, from 2019 July 18 through 2020 July 4. Targets were observed whether or not they were previously known to vary. These observations were made possible by \tess\ Guest Investigator (GI) program \#G022141. All 187 observed systems are highlighted in Figure \ref{fig:varindex}. We download calibrated light curves from the Mikulski Archive for Space Telescopes (MAST)\footnote{\url{https://archive.stsci.edu/tess}}. Light curves were automatically reduced and corrected for instrumental systematics using the \tess\ data processing pipeline\footnote{\url{https://heasarc.gsfc.nasa.gov/docs/tess/pipeline.html}} \citep{jen16}. We use the pre--search data conditioning \textsf{PDCSAP\_FLUX} values, which are simple aperture photometry (\textsf{SAP\_FLUX}) values corrected for systematic trends common to all stars on that chip. For stars that were observed in more than one sector, we combine all observations together into a single light curve, removing any systematic offsets in the average flux between sectors. We also record for each target light curve its associated crowding metric \textsf{CROWDSAP}, which is defined as the ratio of target flux to total flux in the \tess\ aperture. 

In order to identify and properly characterize the nature of photometric variations, we use three different techniques, including (i) discrete Fourier transforms, (ii) phase-folded light curves, and (iii) visual inspection. 

\subsection{Discrete Fourier Transform}
In order to search for periodicities in our light curves, we computed discrete Fourier transforms (DFTs) using \textsf{Period04} \citep{len05}, out to the Nyquist frequency. The frequency resolution, which we define as the inverse of the observation run length, varied from target to target since some were observed in multiple sectors. For the typical 27-d long light curve, the resolution was 0.037 d. The Nyquist frequency for all light curves was set by the 2-min cadence and equal to 360 d$^{-1}$. Before computing any DFTs, we first removed long--term ($>$10 d) trends in the data sets by fitting and subtracting simple polynomial functions up to third order. Otherwise, 1/$f$ noise could hide low-amplitude signals at the longest periods. We note that most $p$--mode oscillations from sdBV$_r$ stars, if present, would be found above the Nyquist frequency and not detectable in our data. 

\begin{figure*}[t]
\label{fig:fourier}
\centering
\includegraphics[width=2.1\columnwidth]{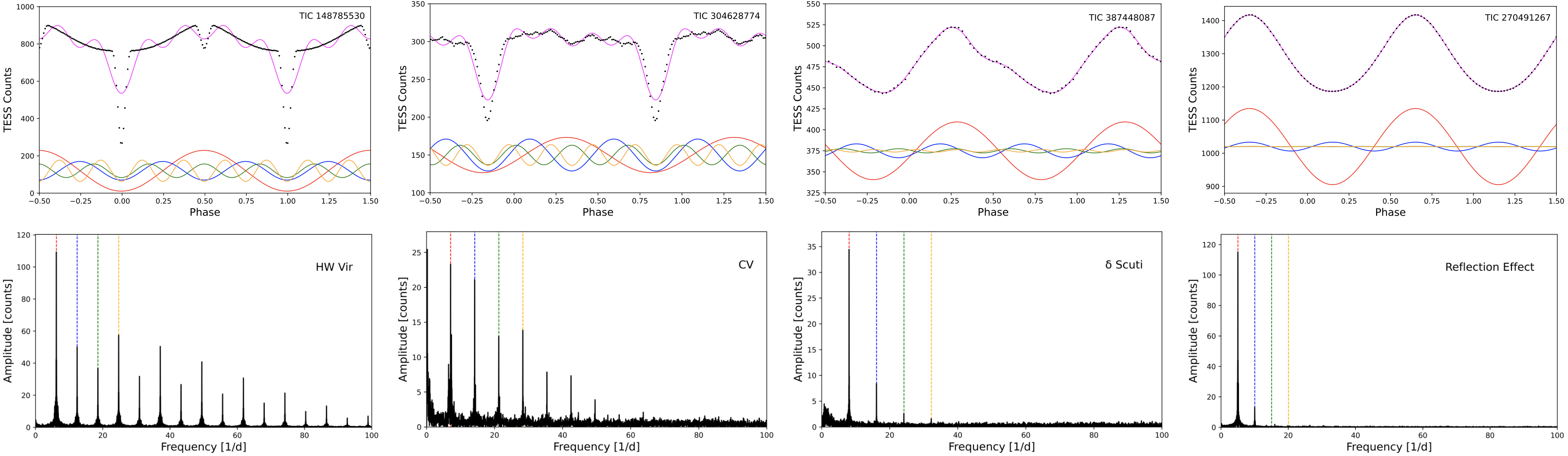}
\caption{Example of a simple Fourier analysis of variable star light curves from \tess\ Cycle 2. {\em Top Panels:} Phase-folded light curves are shown for an example HW Vir, CV, $\delta$ Scuti, and reflection effect system, from left to right. Red, blue, green, and orange lines illustrate the amplitudes and phases of the fundamental, 1st, 2nd, and 3rd harmonics of the DFT, respectively. The purple lines show the superposition of these four sine waves. {\em Bottom panels:} Discrete Fourier transforms of the light curves shown above, which illustrate the pattern of relative power in the fundamental and harmonics needed to generate each light curve shape.}
\end{figure*}

\subsection{Phased-Folded Light Curve}
While the DFT can help quantify the frequency and amplitude of any signals present, it does not reveal information regarding the relative {\em phases} of such signals, which help set the overall shape of the light curve. Once the orbital period of a binary was determined in the DFT, we phase-folded and then binned the entire light curve on this period. For consistency, we took as the number of bins the (rounded) square root of the number of data points in each light curve. If the DFT revealed multiple harmonics, we tried folding the light curve on twice the period of the highest peak (and other factors) to ensure we had found the correct orbital period. This is most important for ellipsoidally modulated systems and contact binaries, for which the first harmonic amplitude may exceed the fundamental amplitude by a significant factor. Our strategy for pulsating stars depended on the nature of the pulsations. If a pulsator exhibited few or no incommensurate frequencies, we phase--folded the data over the strongest mode to investigate its pulse shape. We did not attempt phase--folding for pulsators showing an array of strong incommensurate frequencies (e.g., like sdBV$_s$ stars). Lastly, some light curves showed obvious flux variations but with signals that were incoherent and impossible to fold over.

\subsection{Visual Inspection \& Classification}
Finally, we visually inspected each target's full light curve and phase--folded light curve --- something only made possible by our limited number of targets. After taking into account each target's DFT, phase-folded light curve, and position in the \gaia\ CMD, we assigned it one of the following variability classifications: `HW Vir' (eclipsing sdO/B$+$dM/BD), `reflection' (non--eclipsing sdO/B$+$dM/BD), `ellipsoidal' (sdO/B$+$WD), `CV' (non--eclipsing CV), `CV (eclipsing)' (eclipsing CV), or `BV variable' (B--type variables near the main sequence). Systems showing no significant peaks in the DFT were classified as `NOV' (not observed to vary), while variables with uncertain classifications were labelled as `unclear.' In some cases, additional information provided by a Fourier diagnostic plot (see \S \ref{sec:fourier}) and spectroscopy (see \S \ref{sec:spectroscopy}) was taken into account during the classification process.


\section{Fourier Diagnostic Tool for \\Light Curve Classification}
\label{sec:fourier}

Traditionally, light curves of binary hot subdwarf systems, unlike pulsating stars, have not been subjected to Fourier analysis due to the limited number of orbits for which they are observed or complicated window functions from daytime gaps. Continuous 27$+$ day, 2 min--cadence observations from \tess\ make such an analysis fruitful for a large number of hot subdwarfs and related objects. Consequently, we spent some time exploring the efficacy of using Fourier series of the systems as diagnostic tools for classification.

\begin{figure}[t]
    \centering
\includegraphics[width=1.0\columnwidth]{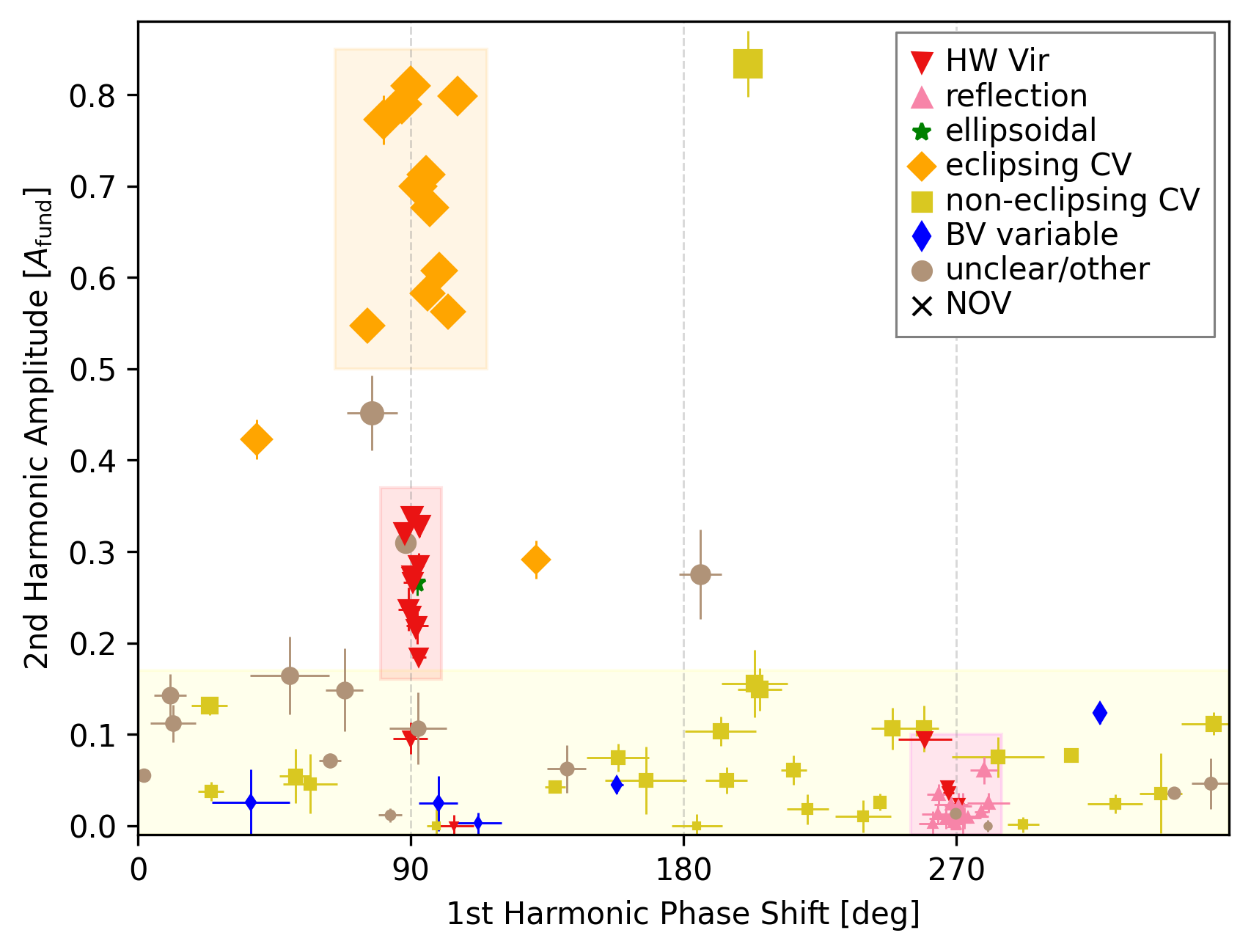}    
\caption{Fourier diagnostic plot for observed variables in our survey. The amplitude of the second harmonic (relative to the fundamental) is plotted against the phase of the first harmonic (relative to the fundamental). Symbol sizes indicate the strength of the first harmonic amplitude (relative to the fundamental). A system's position in this plot helps reveal its light curve morphology. Shaded boxes denote regions where HW Vir binaries (red box), eclipsing CVs (orange box), non-eclipsing CVs (yellow box), and reflection effect systems (pink box) tend to cluster. Only systems with amplitude ratio uncertainties $<$ 0.05 and phase shift uncertainties $<$18$^{\circ}$ are included.}

\label{fig:fourier_diagnostic}
\end{figure}

As can be seen in Figure 2, the pattern of harmonics present in the DFT of a light curve can reveal the light curve shape and help classify the type of photometric variation. 
For example, the DFTs of reflection effect systems are dominated by a fundamental peak and much smaller first harmonic, whereas eclipsing HW Vir binaries and cataclysmic variables show a plethora of harmonics whose relative amplitudes can help distinguish one shape from the another. 

The relative phases of the fundamental signal and its harmonics turn out to be quite important. To quantify this information, we fit the following series of harmonics to each \tess\ light curve from our \varindex\ survey:

\begin{equation}
    f(t) = f_o + \sum_{i=0}^2 a_i \sin \left( \frac{2 \pi t}{P/(i+1)} + \phi_i \right),
\end{equation}

\noindent where $i=0$ is defined as the fundamental ($i=1$ the first harmonic, etc.), $P$ is the fundamental period, $a_i$ the amplitude of each signal, and $\phi_i$ its phase. Perhaps the most important distinguishing factor between light curve shapes is the {\em phase} of the first harmonic ($i=1$) with respect to the phase of the fundamental ($i=0$), which is given by:

\begin{equation}
    \Delta \phi_{1-0} = -2.0 \, \phi_0 + \phi_1
\end{equation}

\noindent Light curves that are roughly symmetric about their peak flux have $\Delta\phi_{1-0}$ = 90$^{\circ}$ or 270$^{\circ}$. For example, the reflection effect in non--eclipsing sdB+dM/BD binaries is symmetric about its peak and deviates from a sinusoidal shape, with the crests being slightly sharper and the troughs being flatter. Although a single sine wave at the orbital period (the fundamental signal) generates the majority of this shape, a low--amplitude first harmonic with $\Delta \phi_{1-0}$=270$^{\circ}$ is also necessary to make the crests sharper and troughs flatter. Consequenly, reflection effect DFTs are dominated by a strong fundamental peak and much weaker first harmonic; no other harmonics are typically present. For the eclipsing HW Vir and CV binaries, a much stronger first harmonic signal with $\Delta \phi_{1-0}$=90$^{\circ}$ is needed to begin fitting the primary eclipses. Strongly asymmetric light curve profiles are produced with $\Delta \phi_{1-0}$ values nearer 0$^{\circ}$ or 180$^{\circ}$. For example, reverse sawtooth shaped light curves of strong radial pulsators like some $\delta$ Scuti or Blue Large--Amplitude pulsators (BLAPs; \citealt{pie17}) tend towards $\Delta \phi_{1-0}$ = 360$^{\circ}$ (=0$^{\circ}$) in order to generate their characteristic fast rise, slow decay profiles. Like the reflection effect systems, they rarely need power at the second harmonic and beyond to recreate their shapes. Any normal sawtooth-shaped light curve profiles (i.e., slow rise, fast decay) would have $\Delta \phi_{1-0}$ = 180$^{\circ}$.
     
\begin{figure*}
\centering
\includegraphics[width=1.6\columnwidth]{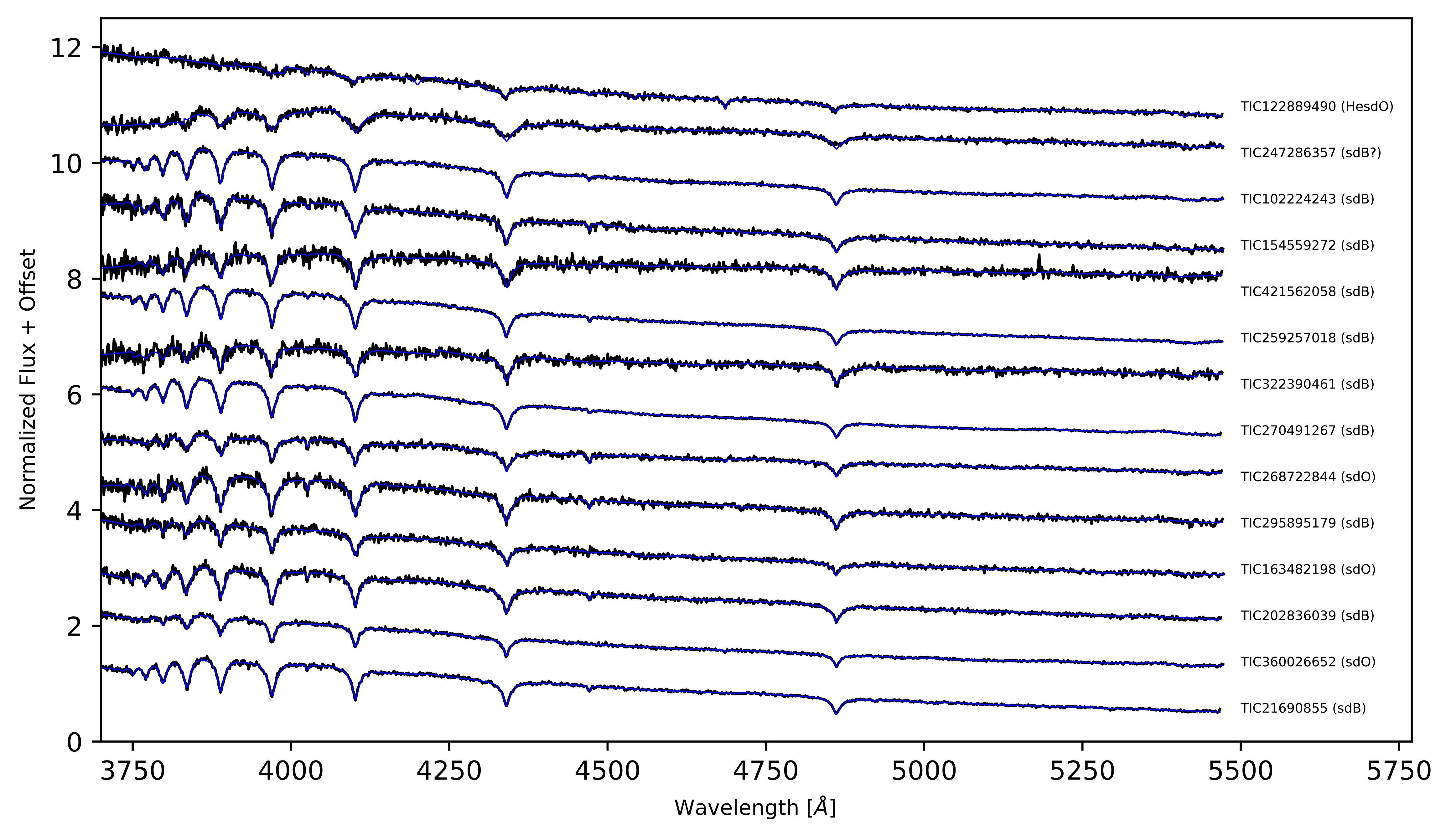} 
\caption{Lick observatory spectra of variable hot subdwarf stars found in our survey (black lines) with their best--fitting {\sc Tlusty/XTgrid} models (blue lines), ordered from shortest (top) to longest (bottom) orbital period. Section \ref{sec:hwvir} and Table \ref{tab:reflection_spec} list the parameters associated with these models. Fluxes shown are normalized by the mean, and offsets have been applied in order to show all spectra together.}
\label{fig:spectra_hot_subdwarfs}
\end{figure*}

While the phase of the first harmonic is a powerful diagnostic for classifying light curve shapes, it alone is not enough. The relative amplitudes of the first harmonic ($A_1$) and second harmonic ($A_2$) can help further separate similar light curve profiles.

Figure \ref{fig:fourier_diagnostic} presents a simple Fourier diagnostic plot that illustrates how different light curves shapes can be separated based upon the amplitude and phases of the fundamental signal and its first few harmonics in the DFT. In the figure, the amplitude of the second harmonic (as a fraction of the fundamental amplitude) is plotted against the phase of the first harmonic (with respect to the fundamental phase). Additionally, the relative strength of the first harmonic is represented through the sizes of the markers. Several classes of objects congregate in distinct parts of this diagram. The non-eclipsing reflection effect binaries (sdB+dM), whose DFTs are dominated primarily by a fundamental and first harmonic, are found with $\Delta \phi_{1-0}$=270$^{\circ}$ (needed to make them symmetric, with pointier crests and flatter troughs) and at low $A_2/A_0$ (i.e., a fundamental and first harmonic can almost reproduce the reflection effect shape by themselves). The eclipsing HW Vir binaries sit in a tight cluster near $\Delta \phi_{1-0}$=90$^{\circ}$ (needed to make the shape symmetric with pointier troughs and flatter crests), with second harmonic amplitudes 20-35\% of the fundamental amplitude (the sharper eclipse shape requires several harmonics). HW Virs with grazing eclipses, however, have light curve shapes dominated by the relection effect, and so they actually appear along with the other reflection effect binaries. Eclipsing CVs are found at the same phase shift as the HW Vir binaries, but with much stronger second harmonic amplitudes ($>$50\% of the fundamental). Non--eclipsing CVs, which can display a diverse array of light curve shapes depending on the activity, disk structure, etc., are not constrained to one particular phase shift, but they do generally show weaker second harmonic amplitudes.

While this diagnostic plot alone may not be sufficient for classifying a system, combining this information with a target's position in the \gaia\ CMD provides a surprisingly effective tool for classification. Most notably, it can help distinguish low--inclination reflection effect systems, which have light curves that barely deviate from a sinusoid, from truly sinusoidal flux variation systems. We are currently exploring the full potential of this methodology and will discuss in a future paper.

\section{Lick Spectroscopy} 
\label{sec:spectroscopy}

\begin{figure*}
\centering
\includegraphics[width=1.524\columnwidth]{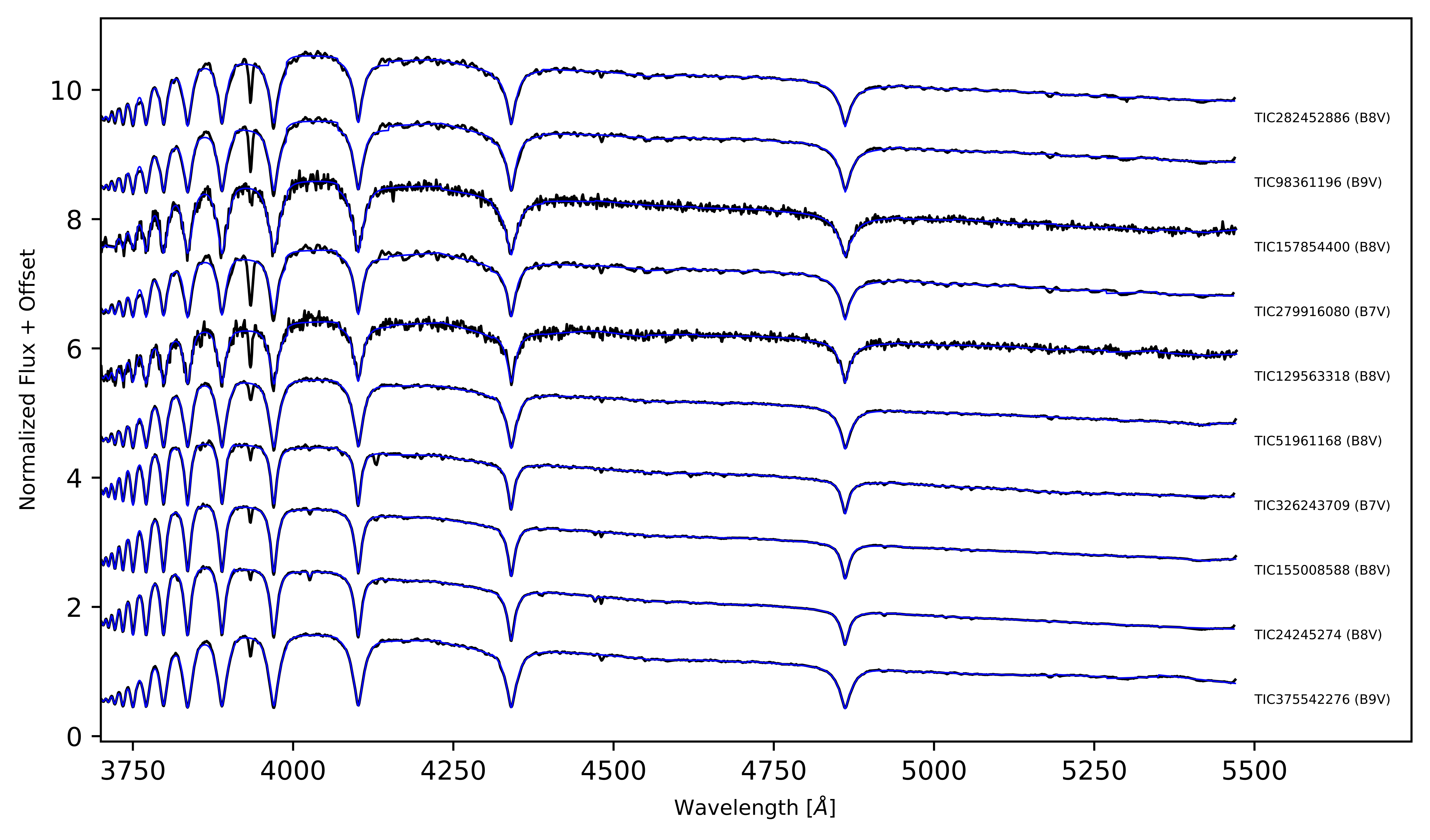} 
\caption{Lick observatory spectra of variable B--type main sequence stars found in our survey (black lines) with their best--fitting {\sc Tlusty/XTgrid} models (blue lines), ordered from shortest (top) to longest (bottom) orbital period. Table 5 lists the parameters associated with these models. Fluxes shown are normalized by the mean, and offsets have been applied in order to show all spectra together.}
\label{fig:spectra_B-type_stars}
\end{figure*}

\begin{figure*}
\centering
\includegraphics[width=1.524\columnwidth]{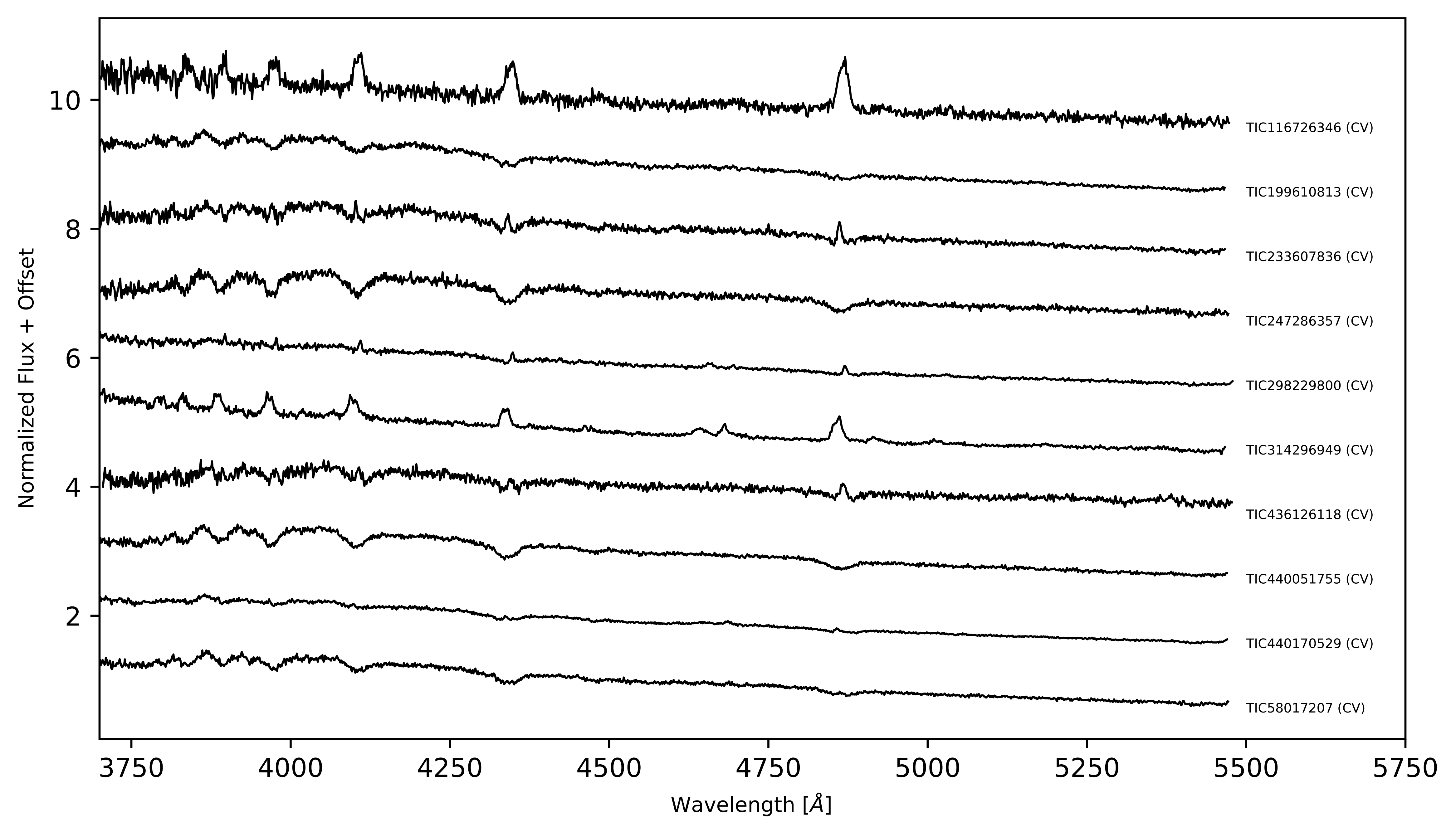} 
\caption{Lick observatory spectra of non-eclipsing cataclysmic variables (or some candidate magnetic white dwarfs) identified in our survey (black lines). 
The emission components of the H Balmer and He lines indicate the presence of a disk and/or extended optically thin gas around these stars, which {\sc XTgrid} is unable to reproduce with the static and  plane--parallel {\sc Tlusty} atmosphere models.
Table 4 gives some details on these systems. Fluxes shown are normalized by the mean, and offsets have been applied in order to show all spectra together.}
\label{fig:spectra_CVs}
\end{figure*}

In order to confirm some of our less certain classifications or measure atmospheric parameters of the primary star in others, we obtained spectroscopy with the Kast spectrograph on the 3--m Shane telescope at Lick observatory for 44 of our targets on the nights of 2020 September $14-15$. We used the 600/4310 grism to cover the spectral range $3350-5470$\,\AA\ with a resolution of $R \sim 1300$. The observed targets had no prior spectral classification in the literature we could find and were selected because they fell in one of the following three categories: (i) candidate HW Vir and reflection effect binaries; (ii) systems with coherent and interesting light curve variations but unclear classification; and (iii) candidate cataclysmic variables.

\begin{table*}[t]
\centering
\label{tab:results}
\begin{tabular}{lllll}
\hline
TIC ID & \gaia\ DR2 ID & Alias & Period & Type \\
 & & & [d]  &\\
\hline
279916080	&	2104025004235555200	& HD177485 &	0.413244	&	B7V variable	\\
326243709	&	1980077058440787712	& TYC3617-2833-1 &	1.497175	&	B7V variable	\\
129563318	&	 1928596794450900480	& J225652.56+390822.3 &	0.766450	&	B8V variable	\\
155008588	&	1935158778628052608	& &	1.677106	&	B8V variable	\\
51961168	&	 426510559099112832	& TYC4017-2123-1 &	1.413124	&	B8V variable	\\
157854400	&	 2097817200244791936	& J280.3949+38.9988 &	0.331006	&	B8V variable	\\
24245274	&	 2045791707334881536	& &	2.049600	&	B8V variable	\\
282452886	&	 1858126238071133184	& HD335181 &	0.046796	&	B8V variable	\\
375542276	&	 1825514757550000768	& HD 231977 &	3.077375	&	B9V variable	\\
98361196	&	 1861191062326013696	& HD334099 &	0.091219	&	B9V variable	\\
159580213	&	 2126637221076540544	& KIC 8751494 &	0.359475	&	blend (CV+EB)	\\
63452125	&	 2128235773544310400	& NSVS 5547844 &	0.254587	&	CV	\\
396737942	&	 1112772429499375488	& BZ Cam &	0.153691	&	CV	\\
260601559	&	 207022199674484608	& ASASSN-V J050221.01+471144.0 &	0.228155	&	CV	\\
279923419	&	 568635875943206272	& HS 0229+8016 &	0.161365	&	CV	\\
378426751	&	 1826189758928713216	& FY Vul &	0.191306	&	CV	\\
142155545	&	 1014839031890550912 & EI Uma	&	0.268100	&	CV	\\
69026620	&	 890152695314991488	& J108.0089+32.4448 &	0.205289	&	CV	\\
\ldots & \ldots & \ldots & \ldots & \ldots \\
\hline
\end{tabular}
\caption{Summary of results from our \tess\ Cycle 2 survey of candidate variable hot subdwarf stars. We include the dominant period seen in the \tess\ light curve for all systems, as well as our own classification. In cases where the true orbital period is {\em not} the dominant signal in the periodogram (e.g., ellipsoidal systems), we include the true orbital period instead of the dominant signal. For variables with incoherent light curves or no clear dominant signal, we assign the period a value of -99. Systems are ordered alphabetically by their classification. Full table is available online.}
\end{table*}

Most of the spectra showed strong H Balmer absorption lines, with some also exhibiting He I and even He II lines. In order to determine a spectral classification for our targets, we utilized the stellar atmopsheric modeling service at Astroserver \citep{nem17}, which uses {\sc XTgrid} to automate the spectral analysis of early type stars with non--LTE {\sc Tlusty/Synspec} models \citep{hub17a,hub17b,hub17c}. The procedure applies an iterative steepest--descent $\chi^2$ minimization method to fit observed spectra. It starts with an initial model and by successive approximations along the $\chi^2$ gradient converges on a best--fitting solution. The models are shifted and compared to the observations by a piecewise normalization, which also reduces systematic effects like blaze function corrections or absolute flux inconsistencies due to vignetting or slit-loss. {\sc XTgrid} calculates the necessary {\sc Tlusty} atmosphere models and synthetic spectra on the fly and includes a recovery method to tolerate convergence failures, as well as to accelerate the converge on a solution with a small number of models. Fitted parameters included effective temperature (T$_{\mathrm{eff}}$), surface gravity acceleration (log $g$), helium abundance (log nHe/nH), rotational velocity ($v \sin i$), and radial velocity shift ($RV$). During parameter determination of hot stars the completeness of the opacity sources included, and departures from LTE are both important for accuracy. We concluded that {\sc Tlusty} models with H$+$He composition delivered sufficient results given the spectral resolution, wavelength coverage, and signal-to-noise of the spectra.

Of the 44 targets we observed, 14 are newly confirmed hot subdwarfs in binaries (Figure \ref{fig:spectra_hot_subdwarfs}), 10 were found to be B--type main sequence stars (Figure \ref{fig:spectra_B-type_stars}), and 10 show either emission lines or multi--peaked absorption lines and are likely cataclysmic variables or magnetic WDs (Figure \ref{fig:spectra_CVs}). Best--fitting atmospheric parameters are presented in Section \ref{sec:hwvir} for the HW Vir binary,  Table \ref{tab:reflection_spec} for the reflection effect systems, and Table 5 for the B--type main sequence stars. Parameter errors are evaluated by mapping the $\chi^2$ statistics around the solution. The parameters are changed in one dimension until the 60\% confidence limit is reached. Since atmospheric parameters for the WDs in the CVs are unreliable due to the emission lines, we do not present their modeling results but only show their observed spectra. \\

\section{Survey Results} \label{sec:results}

\begin{figure*}[t]
\centering
\includegraphics[width=1.5\columnwidth]{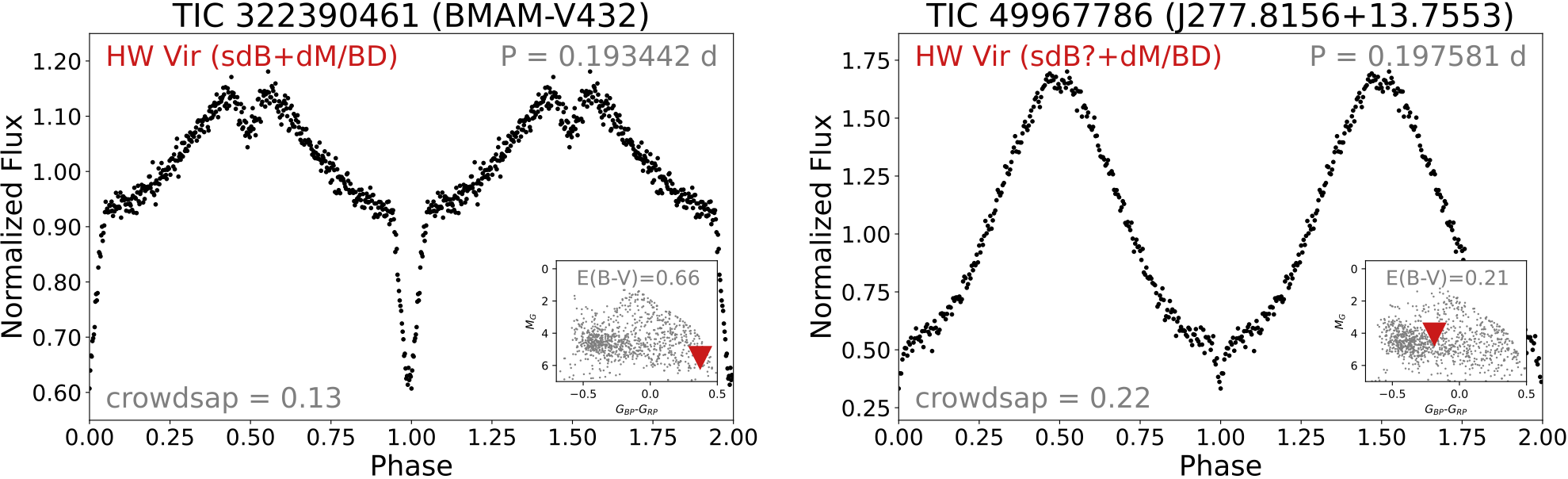}
\caption{Phase-folded light curves for TIC 322390461 and TIC 49967786, two new HW Vir binaries observed by \tess\ and discovered by our \gaia\ variability survey. The inset shows the systems' positions (red points) in the \gaia\ CMD in relation to the candidate hot subdwarfs from \citet{gei19}.}
\label{fig:HW_Vir}
\end{figure*}

We observed a total of 187 known and candidate variable hot subdwarf stars with 2--min cadence photometry from \tess\ during Cycle 2 observations. Figure \ref{fig:varindex} shows their positions in both the Gaia variability plot and \gaia\ CMD, with colors and symbols encoding their classifications. Table 2 summarizes our results for all these objects, including measured periods and light curve classifications.

More than 90\% of our targets show clear photometric variations in their \tess\ light curves, as expected given their anomalously high \gaia\ $G$ flux errors. For the 13 targets (7\% of sample) whose DFTs are consistent with noise, it is currently unclear why their \gaia\ flux errors are so large given their magnitudes. Possible explanations include their having photometric variations with periods much longer than the 27 d \tess\ observations, amplitudes too small to be detected by \tess, strong flux contamination from nearby stars on the same pixel that diminish the observed fractional amplitude, and transient--like events that were caught by \gaia\ but not by \tess.  

The most prominent objects in our sample (71 systems; 38\%) turned out not to be hot subdwarfs at all, but instead cataclysmic variables. They tend to display high \varindex\ values and sit much redward of the ``main sdB clump'' in the \gaia\ CMD (centered at $M_G$=4.5, $G_{BP}-G_{RP}$=-0.4). Reflection effect systems, most of which appear to be sdB+dM/BD binaries, were the second most common member of our sample (26 systems; 14\% of sample). They exhibit much lower \varindex\ values than the average CV and tend to lie closer to the main sdB clump, with the exception of several that show high reddening values.  We also observed 19 eclipsing reflection effect systems (10\% of sample), most of which are also sdB+dM/BD binaries (HW Vir type). Despite their strong primary eclipses, they do not have \varindex\ values as high as one might expect, since any \gaia\ measurements falling deep in the eclipses were likely removed during the iterative sigma--clipping process. We find a handful of ellipsoidally modulated systems (3 targets; $\sim$2\%) with sdB primaries and white dwarf secondaries. At least 10 of our targets (5\% of sample) are B–type main sequence stars, with variations likely due to rotation or binarity. The remaining targets (24\% of sample) are made up of other types of variables/eclipsing binaries and systems with light curves too noisy, strange, or contaminated for us to classify with confidence.

\subsection{HW Vir Binaries}
\label{sec:hwvir}

We obtained high--quality, 2 min--cadence light curves for 19 HW Vir binaries. Their light curves are dominated by three main features: (i) a reflection effect caused by irradiation of a cooler companion by the primary star, (ii) a primary eclipse, when the cool companion blocks the primary, and (iii) a secondary eclipse, when the primary blocks its own irradiated light from the cooler companion. While most of these systems have been previously studied, two are newly confirmed HW Vir binaries and shown in Figure \ref{fig:HW_Vir}. 

TIC 322390461 (Gaia DR2 2219505890166498048) has a period of 0.193443 d and appears to be a typical HW Vir, exhibiting deep primary and secondary eclipses along with a strong reflection effect. Its variability and period were previously reported by AAVSO \citep{wat06}, but the true nature of its variation was unclear until our \tess\ photometry\footnote{We note that \citet{bar21} concurrently discovered this object to have HW Vir--type variability from our \tess\ data.}. The best--fitting model to the Lick spectrum (see Figure \ref{fig:spectra_hot_subdwarfs}) shows the primary to be an sdB with T$_{\mathrm{eff}}$ = 30200 $\pm$ 650 K, log g = 5.96 $\pm$ 0.13, and log(nX/nH) = $-$2.71 $\pm$ 0.34. 

TIC 49967786 (Gaia DR2 4508520908288492672) has an orbital period of 0.197581 d and shows a strong reflection effect with only grazing eclipses. However, its light curve has a \textsf{CROWDSAP} value of 0.22, indicating it contributes just 22\% of the flux in the extracted aperture, so it has significant pixel contamination by other flux sources. Follow--up photometry with higher spatial resolution will be needed to properly characterize this sytem. While we do not possess spectroscopy, its light curve shape, period, and position in the \gaia\ CMD are all consistent with its being a new eclipsing sdB+dM/BD binary. The target was previously reported as a photometric variable by the PanSTARRS1 3$\pi$ survey \citep{ses17}, but the reported period (0.7183115 d) and classification (``RR Lyr'') are inconsistent with our findings.

Our \tess\ data also provide high signal--to--noise light curves for 10 currently unsolved HW Vir binaries uncovered recently by the EREBOS survey \citep{sch19}. Orbital parameters have not been determined yet due to a lack of high S/N photometry and spectroscopy, but many of these \tess\ light curves should be suitable for modeling. Additionally, some known and well--studied HW Vir (e.g., AA Dor) and HW Vir-like (e.g., the PN system UU Sge) binaries also fell in our high \varindex\ sample, as they showed highly anomalous \gaia\ flux errors. While the \tess\ Cycle 2 data add little to our overall understanding of their solved orbital parameters, fits to their eclipses may provide new timing measurements to assist in the search for orbital period and phase changes due to secular evolution of the binary or the presence of circumbinary objects. Light curves of these previously known HW Vir binaries are highlighted in Appendix A.

\begin{table*}
\centering
\resizebox{\textwidth}{!}{\begin{tabular}{llllrlrlrlll}
\hline
\hline
Gaia DR2 ID	&	TIC ID	& Alias &	Period [d]	& \multicolumn{2}{c}{T$_{\rm eff}$ [K]}	 			&	\multicolumn{2}{c}{log (g/cm s$^{-2}$)}	 			&	\multicolumn{2}{c}{log (nHe/nH)}	 		 	&	SpType	&	Remarks	\\
\hline
2118607522015143936	&	122889490	& J278.2669+46.6181 & 	0.070703	&	64110	&	$\pm$ 	2600	&	5.70	&	$\pm$ 	0.30	&	2.30	&	$\pm$ 	0.20	&	He-sdO	&	AAVSO; ATLAS$^{\ddagger}$ (``CBF'')	\\
384468910944036992	&	259257018	& CRTSJ000231.3+425310 & 	0.155784	&	28860	&	$\pm$ 	290	&	5.37	&	$\pm$ 	0.04	&	$-$2.58	&	$\pm$ 	0.12	&	sdB	&	ATLAS (``SINE''), AAVSO$^{\ddagger}$; CSS$^{\ddagger}$	\\
1695662021992833920	&	270491267	& 2MASSJ15292631+7011543 &	0.199643	&	28580	&	$\pm$ 	160	&	5.46	&	$\pm$ 	0.03	&	$-$2.81	&	$\pm$ 	0.13	&	sdBV$_s$	&	g-mode pulsations?; AAVSO$^{\ddagger}$	\\
2208678999172871424	&	268722844	& J231034.22+650033.7 &	0.203795	&	35310	&	$\pm$ 	530	&	5.71	&	$\pm$ 	0.09	&	$-$1.82	&	$\pm$ 	0.17	&	sdOB	&	terrible \textsf{CROWDSAP}; AAVSO	\\
2184734315978100096	&	295895179	& J304.2697+53.7150 &	0.212916	&	26530	&	$\pm$ 	580	&	5.71	&	$\pm$ 	0.10	&	$-$1.99	&	$\pm$ 	0.13	&	sdB	&	brightens every 4 days?; ATLAS; 3$\pi$ $\ddagger$ (``RR Lyr'')	\\
1865490732594104064	&	163482198	& BMAM-V448 &	0.264637	&	36550	&	$\pm$ 	950	&	5.63	&	$\pm$ 	0.11	&	$-$3.48	&	$\pm$ 	0.87	&	sdOB	&	terrible \textsf{CROWDSAP}; AAVSO	\\
391484413605892096	&	202836039	& &	0.280904	&	29410	&	$\pm$ 	470	&	5.48	&	$\pm$ 	0.07	&	$-$2.26	&	$\pm$ 	0.13	&	sdB	&		\\
2073337845177375488	&	360026652	& J194649.77+395937.3 &	0.450977	&	39350	&	$\pm$ 	380	&	5.54	&	$\pm$ 	0.05	&	$-$3.00	&	$\pm$ 	0.20	&	sdO	&	ATLAS (``SINE'') 	\\
2046397439474347904	&	21690855	& J192902.42+332155.2 &	0.580537	&	28610	&	$\pm$ 	120	&	5.40	&	$\pm$ 	0.04	&	$-$2.50	&	$\pm$ 	0.10	&	sdB	&	terrible \textsf{CROWDSAP}	\\
\hline
\multicolumn{12}{l}{$^{\ddagger}$variability detected, but reported period off by an integer value of the true period}\\
\multicolumn{12}{l}{$^{\dagger\dagger}$variability detected, but reported period was incorrect and not an integer value of the true period}\\
\end{tabular}}
\caption{New reflection effect binaries from our \gaia\ \varindex\ survey with spectroscopically confirmed hot subdwarf primaries.}
\label{tab:reflection_spec}
\end{table*}

\begin{figure*}[t]
\centering
\includegraphics[width=2.1\columnwidth]{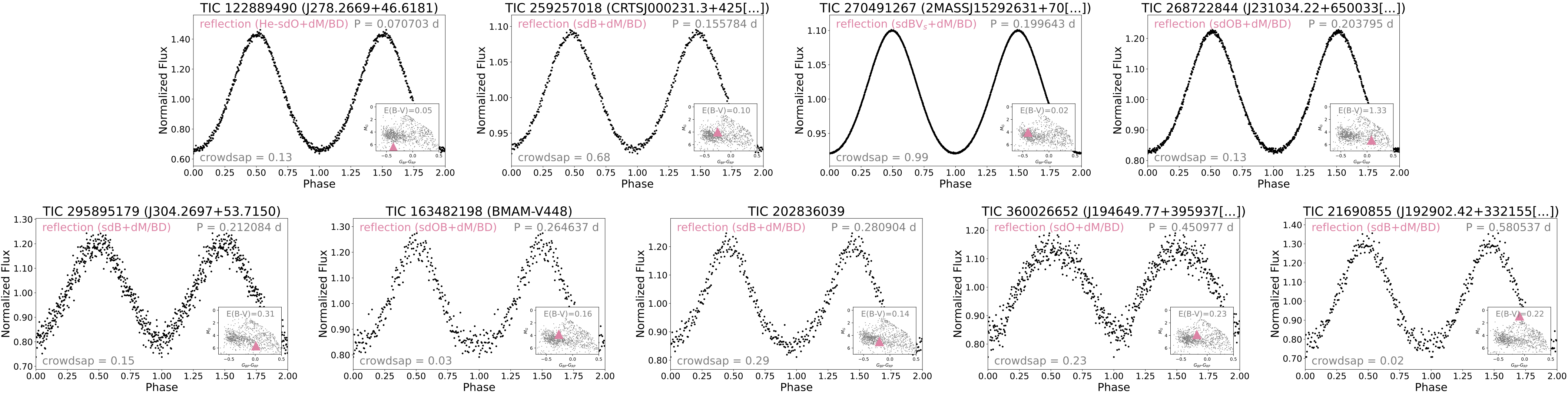}
\caption{\tess\ Cycle 2 phase--folded light curves of new,  spectroscopically confirmed hot subdwarf reflection effect binaries found in our \gaia\ \varindex\ survey.}
\label{fig:reflection}
\end{figure*}

\subsection{Reflection Effect Systems}
\label{sec:reflection}

At least 26 systems in our \gaia\ \varindex\ survey have \tess\ Cycle 2 photometry consistent with non--eclipsing reflection effect binaries. Their light curves exhibit a characteristic, quasi-sinusoidal shape in which the flux peaks are slightly or significantly sharper than the troughs. As the inclination angle decreases, the overall amplitude of the reflection effect decreases, and this difference between the crests and troughs becomes less distinct. Thus, systems that are nearly face--on in our survey, which would have quite sinusoidal light curve shapes, are likely to be recognized as variables but not identified as reflection effect binaries. 

We obtained follow--up spectroscopy for nine of these reflection effect systems, all of which we confirm to have hot subdwarf primaries. Their parameters are summarized in Table \ref{tab:reflection_spec}, and their phase--folded light curves are shown in Figure \ref{fig:reflection}. five are sdB stars, while two show the He II 4686 \AA\ line and are therefore sdOB stars. Two systems are much hotter. Since the hot subwarfs dominate the optical light in all of these systems, we cannot easily classify their much cooler companions, although they are consistent with M dwarfs, brown dwarfs, or perhaps even exoplanets.  We highlight three of the more notable new systems below.

TIC 122889490 (J278.2669+46.6181) exhibits an incredibly strong reflection effect amplitude, with a shape implying the inclination angle is not far from that required for eclipses. Our fits to the spectrum show the primary is a He-sdO with T$_{\rm eff}$ $\approx$ $64{,}000$\,K, log $g$ $\approx$ 5.7, and a high He abundance of log (nHe/nH) $\approx$ 2.3. While photometric variability was previously reported by AAVSO and ATLAS (``CBF''; reported period off by factor of two), its nature was unclear until our \tess\ light curve. The system's 0.070703-d orbital period makes it the fastest known reflection effect system found consisting of a hot subdwarf --- just barely faster than PG 1017-086 \citep{max02}.

TIC 270491267 (2MASSJ15292631+7011543) displays a much weaker reflection effect and contains a cooler sdB primary with T $\approx$ $28{,}500$\,K and log $g$ $\approx$ 5.5. These parameters place the primary in the sdBV$_s$ instability strip. Sure enough, a periodogram of the \tess\ light curve reveals around a dozen signals (see \S\ref{sec:gmode}) that we interpret as $g$--mode pulsations. With $G=12.5$\,mag the system is well--suited for follow-up observations from the ground and an asteroseismological analysis.

TIC 295895179 (J304.2697+53.7150) appears to be a typical sdB+dM/BD reflection system. However, its Sector 16 light curve shows short-duration dips approximately once every 3.7 d. These events last less than one orbital period and only decrease the brightness by roughly half of the reflection effect amplitude. With a \textsf{CROWDSAP} value $<$ 0.2, its light curve is contaminated by nearby stars, and so this extra signal may not come from TIC 295895179. Curiously, its Sector 14 and Sector 15 observations do not show similar brightening events. 

In addition to the newly confirmed hot subdwarf reflection effect systems presented above, we also observed several other known and candidate reflection effect binaries, which are presented in Appendix B. We find six new reflection effect systems with primaries that were previously classifed spectroscopically as sdO/B stars in the literature. We discuss two of them below.

TIC 430960919 (KPD2215+5037) exhibits a quasi--sinusoidal light curve variation with a period of 0.307903 d. The periodogram reveals a first harmonic with amplitude (relative to the fundamental) and phase consistent with a lower--inclination reflection effect. The primary star in this system  was first catalogued as an sdB star by \citet{dow86} in the KPD survey, and \citet{cop11} later reported atmospheric parameters of T$_{\mathrm{eff}}$ = $29{,}600$\,K and log $g$ = 5.640. Additionally, they reported the system as an RV variable with $K = 86.0\pm1.5$\,km~s$^{-1}$ and $P$ = 0.809146\,d. Since then, the companion type has been denoted as either a MS star or WD. However, our  \tess\ observations clearly show KPD 2215+5037 is a non--eclipsing sdB+dM/BD reflection effect system with P = 0.307902752 d. We note that the \citet{cop11} orbital period was determined from two sets of six spectra obtained over three days, with a nearly six--year gap in between the two sets. Thus, their frequency with the smallest $\chi^2$ value was likely an incorrect alias.

TIC 229751806 (HS 1843+6953) was previously observed by \citet{ost10} and \citet{bou17} and found not to vary. However, these searches focused on finding rapid pulsations and would not have been tuned to the relatively long, 0.336711-d period we find from the \tess\ light curve. Past spectroscopic studies by \citet{ede03} show the primary is an sdB with T$_{\mathrm {eff}}$ = 38000 K and log $g$ = 5.60. The \tess\ light curve shows a strong and clear reflection effect, and so we classify HS 1843+6953 as a new sdB+dM/BD reflection effect system. We note that it falls near the northern ecliptic pole and was observed in nearly all Cycle 2 sectors. 

We find seven new candidate hot subdwarf reflection binaries for which classification spectroscopy does not exist, two of which we highlight below.

TIC 138025887 (J088.7202+77.7617) was originally identified as an {\em eclipsing} HW Vir binary in the EREBOS survey, with a period of 0.20371 d \citep{sch19}. However, we do not find any coherent signals when folding our \tess\ photometry on this period. Instead, we find only a clear reflection effect shape when folding on a period of 0.169188 d. The system's position in the \gaia\ CMD, coupled with this light curve shape, imply TIC 138025887 is a non-eclipsing sdO/B+dM/BD binary.
	
TIC 142785398 (NSVS 842188) is a moderately bright ($G=12.7$\,mag) reflection effect system with period 0.194935\,d. Its periodogram shows a strong fundamental ($f$ = 5.1299\,d$^{-1}$) signal and several harmonics. After pre--whitening these frequencies, the periodogram reveals a residual 8$\sigma$ signal at 25.2829 d$^{-1}$ (0.9493\,hr). While this frequency falls close to the position of the fifth harmonic of the binary fundamental ($5f$ = 25.6495\,d$^{-1}$), it is separated from this frequency by roughly 10 resolution elements and thus appears to be {\em incommensurate} with the orbital frequency. The nature of this signal remains unclear, but we note it falls within the range of pulsations frequencies seen in sdBV$_s$ stars. TIC 142785398 falls in the main sdB clump the \gaia\ CMD and is likely a non-eclipsing sdO/B+dM/BD binary.

Finally, we also obtained \tess\ Cycle 2 photometry of several known and well--studied reflection effect systems, as they were flagged for observations due to their large \gaia\ \varindex\ values. Two of the more notable systems are highlighted below.

TIC 384992041 (GALEXJ0321+4727) was identified as a non--eclipsing sdB+dM reflection effect system by \citet{rit03} and \citet{kaw10}. A periodogram analysis of our \tess\ light curve reveals more than a dozen significant periodocities (see \S\ref{sec:gmode}). Since the system's previously reported temperature and surface gravity place it in the sdBV$_s$ instability strip \citep{nem12}, we identify these signals as $g$--mode pulsations and classify GALEXJ0321+4727 as a non--eclipsing sdBV$_s$+dM binary. 

TIC 23709993 (PG 1348+369) was first classified as an sdOB star in the PG survey \citep{gre86}. \citet{ake00} reported quasi--sinusoidal variability with a period of 3.31412601\,d, which was later shown to be a reflection effect through continued monitoring \citep{wat06}. The most recent spectral classifications by \citet{gei20} list the system as an sdO star with dM companion showing emission lines. Our \tess\ light curve shows a strong, high S/N reflection effect with orbital period $P$ = 3.3\,d, consistent with past studies. Due to the system's long period, high flux ($G=13.4$\,mag), and emission lines from the dM, spectroscopic monitoring of PG 1348+369 could lead to measurements of both $K_1$ and $K_2$ and, consquently, the mass ratio. If progress continues towards constraining the inclination angles in reflection effect systems via modeling of high S/N light curves (Schaffenroth et al, in prep), dynamical masses could be determined for PG 1348+369.

\subsection{Ellipsoidally Modulated Binaries}

\begin{figure*}[t]
\centering
\includegraphics[width=2.0\columnwidth]{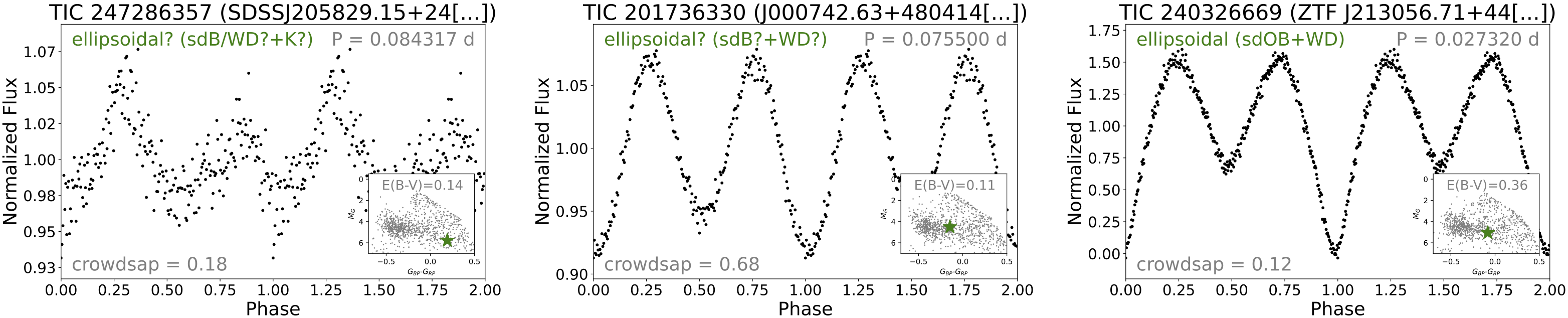}
\caption{\tess\ Cycle 2 phase--folded light curves of four ellipsoidally modulated hot subdwarf binaries. TIC 247286357 and TIC 201736330 represent new candidate ellipsoidal sdB systems, while both TIC 272717401 and TIC 240326669 were previously discovered.}
\label{fig:ellipsoidal}
\end{figure*}

At least three systems with high \gaia\ \varindex\ values observed in \tess\ Cycle 2 showed ellipsoidal modulations, and they are shown in Figure \ref{fig:ellipsoidal}. Two are new systems and discussed below.

\begin{figure*}[t]
\centering
\includegraphics[width=2.0\columnwidth]{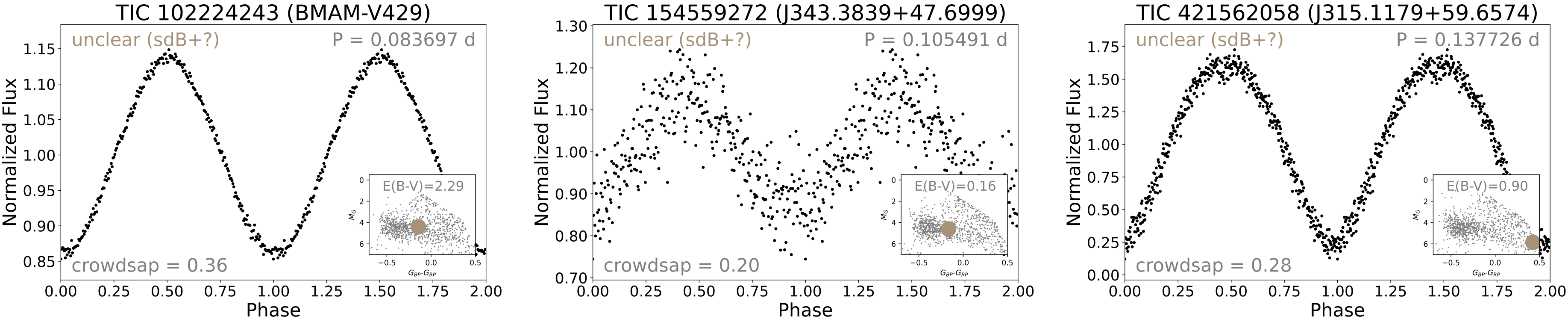}

\caption{\tess\ Cycle 2 phase--folded light curves of new spectroscopically confirmed hot subdwarfs that are variable but of an unknown type.}
\label{fig:other}
\end{figure*}

TIC 247286357 (SDSSJ205829.15+240117.2) appears to show ellipsoidal modulation in its light curve, presumably due to gravitational deformation of one of the two stars. Consecutive troughs are slightly uneven --- a gravity darkening effect often seen in such systems. Similarly, consecutive crests are uneven, which could be a sign of Doppler beaming. Model fits to our observed Lick spectrum indicate T$_{\rm eff}$ $\approx$ $38{,}000$\,K and log $g$ $\approx$ 6.7 but do {\em not} fit the cores of the H Balmer lines well. It is possible the absorption features are contaminated with H emission lines, which would imply TIC 247286357 is a mass--transferring system. Alternatively, the absorption features could be rotationally broadened.  \citet{per16} previously identified TIC 247286357 as a system with infrared excess and tentatively classified the secondary as a K--type main sequence star. We also note that the object sits at a redder color and lower luminosity than the main sdB clump in the \gaia\ CMD. A spectrum with higher S/N and resolution will be necessary to determine whether the primary is a hot subdwarf, low--mass He core pre--WD, or some other object --- and to confirm the nature of the cooler companion. 

TIC 201736330 (J000742.64+480414.4) shows a clear, quasi-sinusoidal light curve shape with an orbital period of 1.8 hr. While the two crests reach approximately the same brightness, the two troughs are cleary uneven --- a potential indication of tidal deformation and strong gravity darkening in the primary component. The system lies within the main sdB clump in the \gaia\ CMD and is consistent with an sdB$+$WD binary. At such a short orbital period and relatively bright magnitude ($G=15.1$\,mag), it should be straightforward to obtain time--series spectroscopy and measure the sdB's RV semi--amplitude, which would exceed 100\,km\,s$^{-1}$ if truly an ellipsoidal system. We note that TIC 201736330 was previously identified as a variable star by AAVSO but at half the true orbital period \citep{wat06}.

The other ellipsoidal system observed in our high--\varindex\ sample has been studied previously:

TIC 240326669 (ZTF J213056.71+442046.5) represents the first Roche lobe--filling hot subdwarf binary found, and it was discovered by \citet{kup20a} in the Zwicky Transient Factory survey only shortly before our own independent \tess\ observations. This rapid, 39--min binary consists of an sdOB transferring mass onto a WD. It, along with the He--sdOB+WD binary ZTF J205515.98+465106.5 \citep{kup20b}, represent the only two members of this newly discovered class of Roche lobe--filling hot subdwarfs. Through the emission of gravitational waves, these binaries should eventually come into contact and either explode as an underluminous thermonuclear supernova, or evolve into a single, massive white dwarf.

\subsection{Other New Hot Subdwarf Variables}

Three of our new, spectroscopically confirmed hot subdwarf targets show strong light curve variations that are unclear in nature. Their folded light curves are shown in Figure \ref{fig:other}. 

TIC 102224243 (BMAM-V429) shows a strong, sinusoidal flux variation with period just over two hours. The periodogram reveals only a single peak --- no other harmonics or incommensurate frequencies are present in the data. Modeling its Lick spectrum, we find   $T_{\mathrm{eff}}$ = $26{,}880\pm350$\,K, log $g$ = $5.51\pm0.05$, and log(nX/nH) = $-2.65\pm0.13$. Its atmospheric parameters and position in the \gaia\ CMD are consistent with it being a core He--fusing sdB star with a significantly fainter companion (if it has a companion at all). The light curve variation seems inconsistent with a reflection effect, which only begins to exhibit a purely sinusoidal shape as the inclination angle approaches zero (face--on). However, such nearly face--on systems would also have extremely low reflection effect amplitudes, and the photometric variation seen in BMAM-V429, nearly 30\% peak-to-peak, is larger than that of most high--inclination reflection effect systems. We note that our single Lick spectrum shows rotational broadening of $v \sin i$ $\approx$ 90 km s$^{-1}$. However, due to  the presence of a nearby star (5 arcsec away) with comparable flux, there might be some level of contamination in the spectrum (and \tess\ light curve) affecting these results.  Follow--up time-series spectroscopy is necessary to shed further light on the nature of this system. 

TIC 154559272 (J343.3839+47.6999) exhibits sinusoidal flux variations with a period just over 2.5 hr, but the poor S/N in the folded light curve might hide any additional features. The Lick spectrum is consistent with that of an sdB with $T_{\mathrm{eff}}$ = $27{,}810\pm620$\,K, log g = $5.708\pm0.08$, and log(nX/nH) = $-2.26\pm0.18$ and shows no features indicative of a companion. We note that the single spectrum shows a radial velocity offset of $-126$ $\pm$ 12 km s$^{-1}$, which could be a sign of an RV--variable system. All this is consistent with TIC 154559272 being a non--eclipsing (or grazing eclipsing) sdB+dM/BD binary. However, a higher S/N light curve and radial velocity curve are be needed to know with certainty. 

TIC 421562058 (J315.1179+59.6574) displays a strange light curve shape almost resembling that of an upside--down reflection effect light curve, with flatter crests and sharper troughs. The period is 0.137726 d. While the system has a \gaia\ color much redder than typical sdBs, there is a significant amount of reddening in this location. The Lick spectrum shows strong H Balmer lines and is fitted well with $T_{\mathrm{eff}}$ = $29{,}440\pm1000$\,K, log g = $5.76\pm0.08$, and log(nX/nH) = $-2.40\pm0.50$. We note that the star was previously flagged as a variable in the ATLAS survey, with class 'NSINE' and the same period we detect \citep{hei18}. Variability was also reported by the PS1 3$\pi$ survey, but at a different period (0.5079379822 d; \citealt{ses17}).

\subsection{Candidate g--Mode sdBV$_s$ Pulsators}
\label{sec:gmode}
\begin{figure}[t]
\centering
\includegraphics[width=1.0\columnwidth]{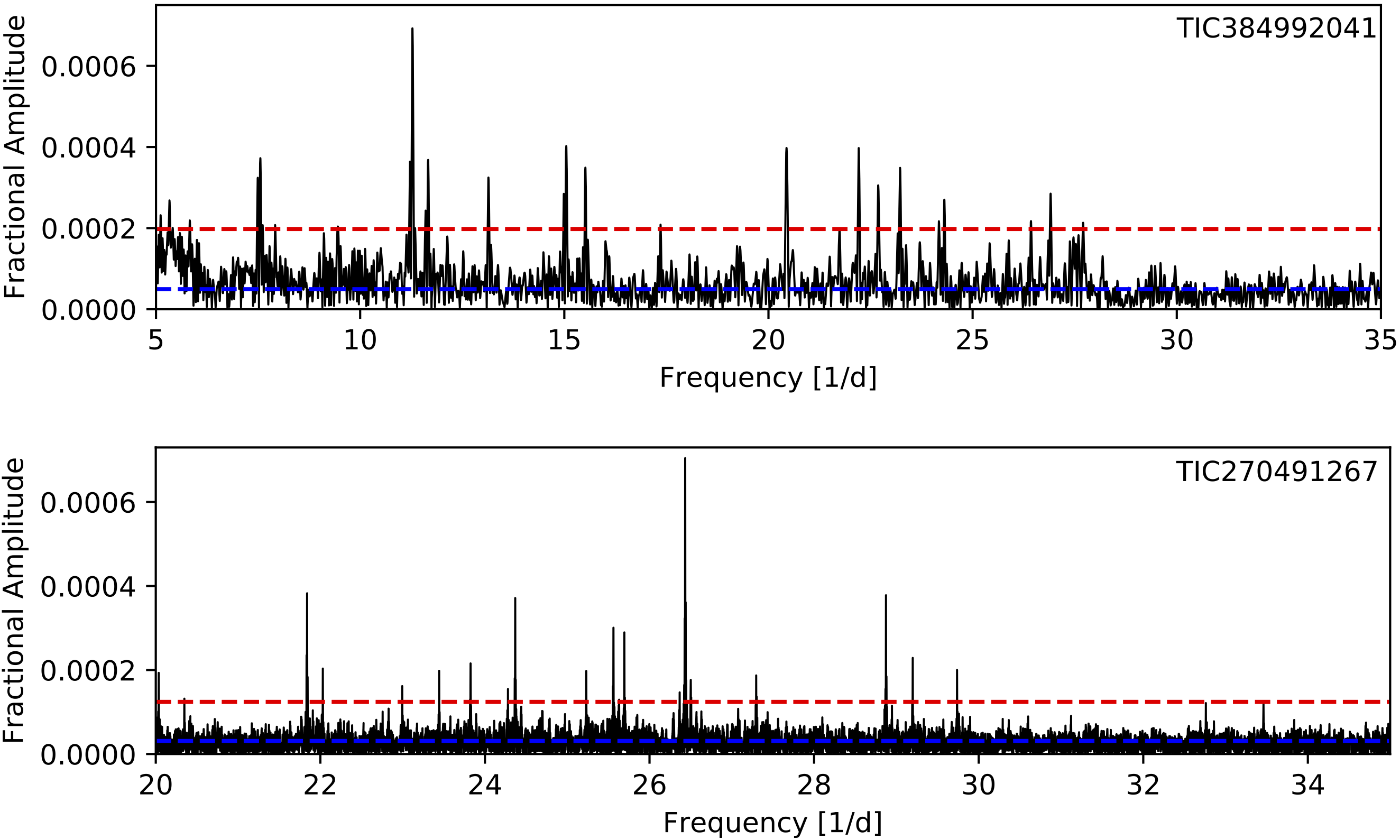}

\caption{Discrete Fourier transforms showing $g$--mode pulsations in the reflection effect binaries TIC 384992041 and TIC 270491267 (after pre--whitening the reflection effect signals). The blue dashed line marks the mean noise level, and the red dashed line marks four times this value.}
\label{fig:pulsations}
\end{figure}

We find what look like candidate g--mode pulsations in two of our targets, both of which are also in reflection effect binaries. Amplitude spectra highlighting their pulsation frequencies are shown in Figure \ref{fig:pulsations}.

TIC 384992041 (GALEXJ032139.8+472716) was flagged as a high \varindex\ object because of its strong reflection effect (see Section \ref{sec:reflection}). However, after pre--whitening the fundamental binary signal and several of its harmonics, the DFT reveals more than a dozen incommensurate frequencies with periods from $0.85-2.15$\,hr and amplitudes all $<$1\,ppt. As noted in \S \ref{sec:reflection}, the system's atmospheric parameters place the star in the sdBV$_s$ instability strip, and so we interpret these oscillations as $g$--mode pulsations. 

TIC 270491267 (2MASSJ15292631+7011543) was also flagged as a high \varindex\ object due to its strong reflection (Section \ref{sec:reflection}). After pre--whitening the light curve of all signals due to the reflection effect oscillation, the DFT reveals more than two dozen significant peaks with periods from $0.8-3.0$\,hr and amplitudes from $0.1-1.0$\,ppt. Atmospheric parameters determined from our Lick spectroscopy also place this star safely in the sdBV$_s$ instability strip, and so we again interpret these oscillations as $g$--mode pulsations. As this system is quite bright ($G=12.5$\,mag), it is especially well--suited for follow-up observations from the ground and a thorough asteroseismological analysis.

\subsection{Cataclysmic Variables / Magnetic WDs}

\begin{table*}
\label{tab:cv_spec}
\centering
\resizebox{\textwidth}{!}{\begin{tabular}{llllll}
\hline
\hline
Gaia DR2 ID	&	TIC ID	&	Alias	&	Dominant Period [d]	&	Emission Components	&	Remarks	\\
\hline
1985702915838387712	&	298229800	&	OR And	&	0.125246	&	He II, C III	& nova--like (NL) \citep{dow01}; AAVSO$^{\dagger\dagger}$	\\
1818529770639712384	&	314296949	&	ASASSN-V J203515.27+220631.0	&	0.131393	&	He II, C III	&  ATLAS$^{\dagger\dagger}$	\\
1866995727800162048	&	116726346	&	NADA-V98	&	0.137981	&	H I (strong)	&	long-period (2.96 d) {\em and} short-period (0.14 d) signals?; AAVSO\\
385980056534991360 & 440051755 & J003.4753+44.8905 & 0.155858 & H I & magnetic WD instead of emission? \\
1832475082783769856	&	436126118	&	ASASSN-V J202511.39+250535.3	&	0.177953	&	H I	& Gaia color blue look like sdB	\\
1624196965939218432	&	199610813	&	2MASSJ16360160+5944411	&	0.218491	&	H I	&	magnetic WD instead of emission?	\\
1988243788429717504	&	249750520	&	J339.3218+50.3868	&	0.219096	&	H I	& small--amplitude, long--period sinusoidal variations; incoherent?; ATLAS\\
382202375099797888	&	440170529	&		&	0.227086	&	H I	&	novae?  ATLAS; \citet{del18} report UV variability\\
2864236144068661376	&	58017207	&	FBS0019+348	&	0.230656	&	H I	&	3$\pi$ survey $^{\dagger\dagger}$	\\

2156722985944854784	&	233607836	&	ASASSN-V J185308.02+611324.1	&	{\em incoherent}	&	H I (strong)	& incoherent but strong signals; AAVSO	\\
\hline
\multicolumn{6}{l}{$^{\ddagger}$variability detected, but reported period off by an integer value of the true period}\\
\multicolumn{6}{l}{$^{\dagger\dagger}$variability detected, but reported period was incorrect and not an integer value of the true period}
\end{tabular}}
\caption{New non--eclipsing cataclysmic variables found in our \tess\ survey with spectroscopy confirming emission components.}
\end{table*}

\begin{figure*}[ht]
\centering
\includegraphics[width=2.1\columnwidth]{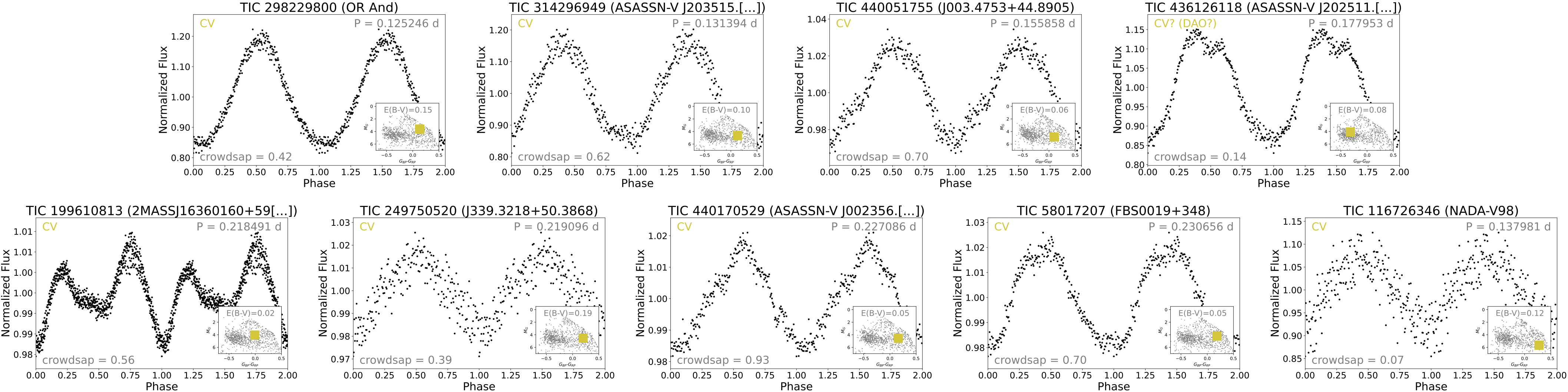}
\caption{\tess\ Cycle 2 phase--folded light curves of new,  spectroscopically confirmed cataclysmic variables found in our \gaia\ \varindex\ survey, further detailed in Table~\ref{tab:cv_spec}. One confirmed CV, TIC 233607836, shows strong but highly incoherent photometric variations, and so we do not show its folded light curve here.}
\label{fig:CVs_confirmed}
\end{figure*}

Approximately one in three variables in our survey turned out not to be hot subdwarf stars, but instead cataclysmic variables (CVs). This should come as little surprise: \citet{abr20} show that nearly all types of CVs fall within the same part of the \gaia\ CMD \citet{gei19} used for their hot subdwarf selection criteria. This includes nova--like (NL) CVs, dwarf novae (DN), classical novae, and intermediate polars. Only WZ Sge and polars are (generally) outside these bounds. CVs tend to exhibit large photometric variations and thus would be expected to show inflated \gaia\ flux errors.

In total, we identify at least 71 CVs in our \tess\ Cycle 2 sample. As seen in Figure \ref{fig:varindex}, they have the highest average \varindex\ values of all types of variables in our survey, and they lie redward of the main sdB clump in the \gaia\ CMD. At least 17 of the observed CVs --- those with the highest \varindex\ values --- show clear eclipses of an accretion disk, hot spots, and other elements associated with eclipsing CVs (see Appendix C). Many of these have been previously identified and studied, but several are new discoveries. Although we have no confirming spectroscopy for those that are new, their light curve morphology, periods, and positions in the \gaia\ CMD make their classification as eclipsing CVs sound.

The remaining 54 systems are either previously known or candidate {\em non--eclipsing} CVs based off of their light curve shapes and positions in the \gaia\ CMD. Since the light curve morphologies of non--eclipsing CVs vary wildy from system to system, it is more difficult to confirm their nature without spectroscopy. We obtained identification spectra for 10 candidate non--eclipsing CVs (see Figure \ref{fig:spectra_CVs}), and their light curves are shown in Figure \ref{fig:CVs_confirmed}. All show either H I, He II, or C III emission lines consistent with CVs (see Table 4).  Folded light curves for other known and potentially new non--eclipsing CVs and magnetic WDs from our survey are shown in Appendix D.

One system worth pointing out is TIC 440051755 (J003.4753+44.8905), which was originally reported as an eclipsing HW Vir binary in the EREBOS project, with an orbital period of 0.797432 d. \citep{sch19}. Our \tess\ photometry reveals a different story and shows a quasi--sinusoidal light curve shape with period 0.155858 d. The Lick spectrum shows broad H Balmer lines that could not be fitted well with any meaningful hot subdwarf or white dwarf model, and thus we cannot report $T_{\rm{eff}}$ or log g values with confidence. Fits to the profiless leaned towards a hot subdwarf with rotational broadening in excess of break--up velocity ($\approx$ 1400 km s$^{-1}$.) A follow--up high resolution spectrum with Keck/ESI shows that the broadening was not real but a consequence of the H Balmer lines being filled in by weak emission lines. Thus, we tentatively classifiy TIC 440051755 as a non-eclipsing CV. We note that for TIC 199610813 and TIC 440051755, their multi--peaked absorption features might be explained better by the presence of a magnetic field combined with gas features around a recent, post--merger sdB, similar to the recently discovered system J22564-5910 \citep{vos21}. Higher resolution spectra at different epochs will shed light on this hypothesis.

\subsection{B--type Main Sequence Variables}

\begin{table*}
\label{tab:B-type_spec}
\centering \resizebox{\textwidth}{!}{\begin{tabular}{llllrlrlrll}
\hline
\hline
Gaia DR2 ID	&	TIC ID	& Alias &	Period [d]	&	\multicolumn{2}{c}{T$_{\rm eff}$ [K]}	 			&	\multicolumn{2}{c}{log (g/cm s$^{-2}$)}	 			&	\multicolumn{2}{c}{log (nHe/nH)}	 		 	&	SpType			\\
\hline
1858126238071133184	&	282452886	& HD335181 &	0.046796	&	12380	&	$\pm$ 	170	&	4.54	&	$\pm$ 	0.04	&	$-$3.1	&	$\pm$ 	1.2	&	B8V		\\ 
1861191062326013696	&	98361196	& HD334099 &	0.091219	&	11700	&	$\pm$ 	180	&	4.41	&	$\pm$ 	0.04	&	$-$3.3	&	$\pm$ 	0.9	&	B9V		\\ 
2097817200244791936	&	157854400	& J280.3949+38.9988 &	0.331006	&	12080	&	$\pm$ 	270	&	4.89	&	$\pm$ 	0.10	&	$-$3.0	&	$\pm$ 	1.4	&	B8V		\\ 
2104025004235555200	&	279916080	& HD177485 &	0.413244	&	13080	&	$\pm$ 	200	&	4.48	&	$\pm$ 	0.05	&	$-$2.7	&	$\pm$ 	0.7	&	B7V		\\ 
1928596794450900480	&	129563318	& J225652.56+390822.3 &	0.766450	&	12600	&	$\pm$ 	320	&	4.54	&	$\pm$ 	0.10	&	$-$3.0	&	$\pm$ 	1.3	&	B8V	 	\\
426510559099112832	&	51961168	& TYC4017-2123-1 &	1.413124	&	11860	&	$\pm$ 	160	&	4.58	&	$\pm$ 	0.04	&	$-$2.5	&	$\pm$ 	0.3	&	B8V			\\
1980077058440787712	&	326243709	& TYC3617-2833-1 &	1.497175	&	13630	&	$\pm$ 	130	&	3.99	&	$\pm$ 	0.02	&	$-$2.7	&	$\pm$ 	0.7	&	B7V		\\
1935158778628052608	&	155008588	& &	1.677106	&	12330	&	$\pm$ 	90	&	4.01	&	$\pm$ 	0.01	&	$-$2.0	&	$\pm$ 	0.8	&	B8V		\\
2045791707334881536	&	24245274	& & 	2.049600	&	12760	&	$\pm$ 	160	&	4.18	&	$\pm$ 	0.04	&	$-$1.3	&	$\pm$ 	0.2	&	B8V		\\
1825514757550000768	&	375542276	& HD 231977 &	3.077375	&	11510	&	$\pm$ 	160	&	4.68	&	$\pm$ 	0.05	&	$-$3.4	&	$\pm$ 	0.9	&	B9V		\\
\hline
\end{tabular}}
\caption{B--type main sequence star variables uncovered in our \gaia\ \varindex\ survey.}
\end{table*}

\begin{figure*}[ht]
\centering
\includegraphics[width=2.1\columnwidth]{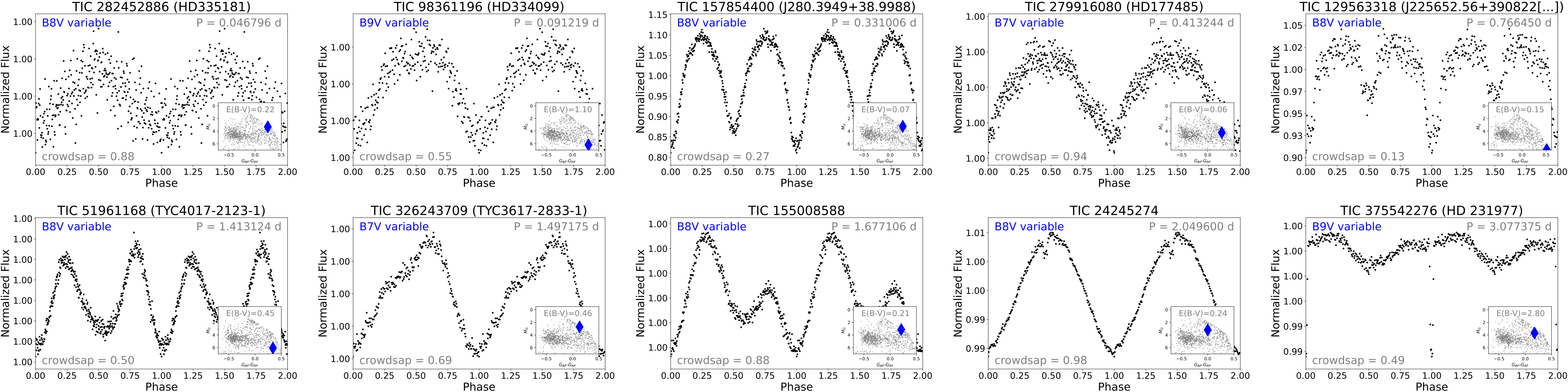}
\caption{\tess\ Cycle 2 phase--folded light curves of spectroscopically confirmed B--type main sequence variables found in our \gaia\ \varindex\ survey. Many are likely variable due to rotation.}
\label{fig:BV}
\end{figure*}

Several dozen systems showed light curve variations with strange shapes and had \gaia\ colors that made them difficult to classify. Follow--up spectroscopy (Figure \ref{fig:spectra_B-type_stars}) shows that at least 10 of these variables are late B--type main sequence stars. We present their atmospheric parameters in Table 5 and their phase--folded light curves in Figure \ref{fig:BV}. We note that all seem to have low He abundances. Since normal B--type main sequence stars show solar--level He abundances, the variables found here could be B$_p$ stars, which are known to be magnetic and show various kinds of photometric modulations due to spots and even flares \citep{bal19,bal20}. While some of our observed variations might be due to rotation, others could be due to binarity. Additionally, it is possible that a few of these objects are slowly pulsating B (SPB) stars, although such stars tend to have higher luminosities in the Gaia CMD than the systems presented here (see Fig. 3 of \citealt{gai19variables}). We also note that many of our B--type variables show significant changes in their light curve amplitudes and morphologies from cycle to cycle, and as such the phase--folded light curves shown in Figure \ref{fig:BV} do not necessarily represent their time--dependent variability accurately. Details on select targets follow.

TIC 282452886 (TDSC 56434) displays several incommensurate frequencies in its DFT. The light curve shown in Figure \ref{fig:BV} is simply folded over the strongest signal (P = 0.046796 d). We note that the target is a known double star made up of two nearly equally bright components separated by $<$1 arcsec \citep{fab02}, and so it isn't clear whether one of the stars display all of these incommensurate periodicites or we are seeing variability from both. We find a B8V spectral classification from modeling our Lick spectrum.  

TIC 98361196 (TDSC 54616) also displays several incommensurate frequencies in the DFT of its light curve, and we simply phase--fold over the strongest signal in Figure \ref{fig:BV}. Like the previous target, it is a double star with comparably bright components separated by $<$1 arcsec \citep{fab02}. We find a spectral classification of B9V. 


TIC 279916080 (TDSC 49544) displays a complicated light curve behavior with significant amplitude/frequency modulation over the \tess\ observation. Its DFT reveals a large number of peaks with periods between 21 min and 1 day. The two dominant signals, around 9.69 hr and 1.79 hr, appear to be heavily amplitude--modulated with modulation periods around 21.3 d and 3.2 d, respectively. The DFT also shows several weaker peaks at higher frequencies with periods as short as $\sim$20 min. Once again, this system is a known double star \citep{fab02}. Thus, it is both unclear from which object these signals originate, and what the true nature of these oscillations might be. The Lick spectrum is fitted well with a B7V model.

TIC 375542276 (HD 231977) shows a strong sawtooth--like variation with period slightly longer than 3 days. Additionally, the folded light curve shows a transit or eclipse--like feature approximately 0.8\% deep. If we assume the B9V has a radius of $\sim$2.7 R$_{\odot}$, then the occulting object would have a radius around $\sim$0.24R$_{\odot}$. We note that in a study of the rotation periods in TESS objects of interest, \citet{can20} report a rotational period of P=3.078 d. We also note that HD231977 is a known visual double with separation $<$1 arcsec (stars are 9.7 mag and 12 mag).

\subsection{Other/Unknown}
In addition to the previous classes of systems discussed, we also observed a handful of other light curve morphologies (e.g., non--hot subdwarf eclpising binaries, unclear sinusoidal oscillations, etc.) in our selection criteria. We were unable to assign classifications to these systems due to the unclear nature of their light curve morphologies, lack of constraints from spectroscopy, and poor S/N in their light curves. These systems are labelled as `unclear' in Table 2. One of these, TIC 402956585 (J305.5398+21.9490), was identified as an eclipsing HW Vir binary in the EREBOS survey using OGLE data \citep{sch19} and found to have an orbital period of 1.267 d; however, the claim of a primary eclipse was based upon only three low--flux measurements. Our \tess\ data show a different light curve shape with much longer period (4--5 d). The eclipse depth reported by \citet{sch19} was 30\%,  and we see no variations larger than 3\%. We note that the \textsf{CROWDSAP} value for TIC 40295658's light curve is 0.65 and indicates smoe level of contamination by background stars, whose flux contributions could limit our ability to detect the HW Vir variations, if they are truly there. Follow--up photometry from an instrument with better spatial resolution is needed to resolve this disagreement. The majority of the other UNCLEAR systems in our sample have nearly sinusoidal light curve shapes -- primarily due to a poor S/N.

\begin{figure}[t]
\centering
\includegraphics[width=1.0\columnwidth]{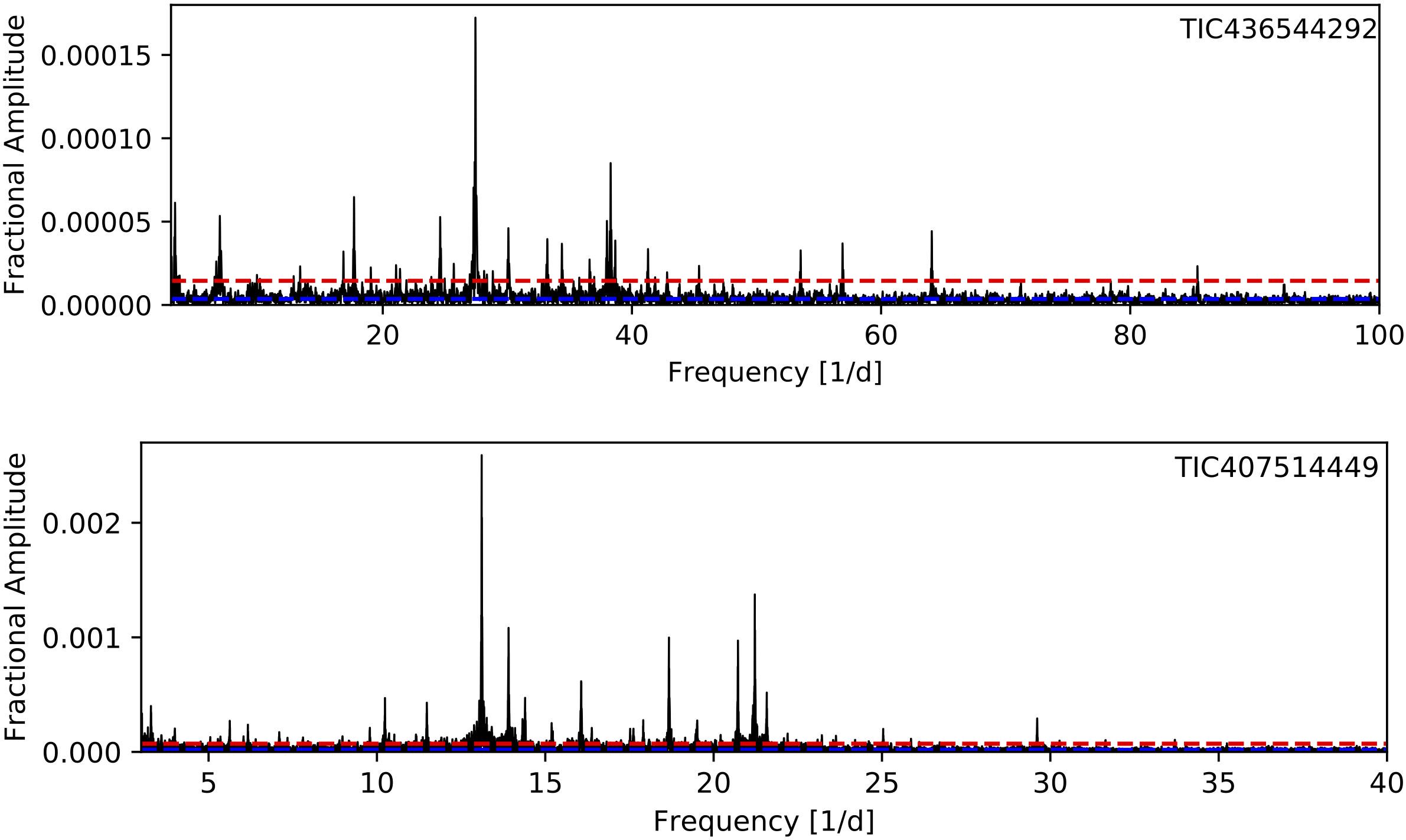}
\caption{Discrete Fourier transforms for TIC 436544292, and TIC 407514449, two candidate $\delta Scuti$ stars. The blue dashed line marks the mean noise level, and the red dashed line marks four times this value.}
\label{fig:scuti}
\end{figure}

TIC 436544292 (HD 5267; 66 Psc), which shows no indication of binarity in its light curve, has a DFT revealing nearly three dozen incommensurate frequencies with periods ranging from 15 min up to 7 hr (see Figure \ref{fig:scuti}). The strongest of these has an amplitude of nearly $\sim$200 ppm while the weakest signals push down to $\sim$10ppm. At $G=6.2$\,mag, it is the brightest object in our survey, and so even these relatively weak signals stick out so high above the mean noise level that they are likely the source of the high \varindex\ value. There is no prior mention of variability at the periods or amplitudes we observe in the literature. Curiously, the reported \gaia\ $G_{BP}$-$G_{RP}$ color is $-$1.839 --- an {\em incredibly} blue value. Although saturation effects can lead to inaccurate colors for the brightest stars observed by \gaia, TIC 436544292 is approximately two magnitudes fainter than the limit where this becomes an issue ($G<4$\,mag; Fig. B.1 of \citealt{eva18}). The amplitudes and frequencies observed in the DFT are consistent with $\delta$ Scuti pulsations. Past studies of TIC 436544292 reveal it to be the more luminous member of a visual binary with an orbital period of 335 yr and either a B9V or A1V spectral type \citep{jah29,osa59,mal12}. 

TIC 407514449 (HD 431) shows no indication of binarity in its light curve but instead approximately 20 incommensurate signals in its DFT with periods from 45 min to 7 hr, and amplitudes from $\sim$300 ppm up to $\sim$ 3 ppt (see Figure \ref{fig:scuti}). Similar to TIC 436544292, this system is both bright ($G$=6.6 mag) and shows an anonamously blue \gaia\ $G_{BP}$-$G_{RP}$ color of -3.142. It is also a visual double consisting of two comparably bright, late A--type stars orbiting once every 531.5 d \citep{par1912,ski14,izm19}. Given the combination of spectral type, frequencies, and amplitudes, it seems likely the observed photometric variations are due to $\delta$ Scuti pulsations.


\subsection{Not--Observed--to--Vary Systems}
Despite their anomalously high \gaia\ flux errors, 13 of our targets showed no statistically significant variations in the DFTs of their light curves. We classify these systems as not--observed--to--vary (NOV) targets. As seen in Figure \ref{fig:varindex}, they are not simply the faintest targets or those with the smallest anomalous \varindex\ values. Many are relatively bright (with $G<15$\,mag) and have \varindex\ values comparable to the rotating B stars, reflection effect systems, and HW Virs. It is currently unclear why they show no photometric variations from \tess\ despite their anomalously high \gaia\ flux errors. Possible explanations include having longer periods than \tess ' observing baseline, amplitudes high enough for \gaia\ to detect in its photometric scatter but too small for \tess\ to have observed, transient--like events that \gaia\ observed but \tess\ did not, noise conspiring in the \gaia\ measurements to inflate its measured scatter and generate a false positive, and contamination from background stars in the \tess\ pixel greatly reducing the fractional amplitude that could be observed. Regarding the last possibility, we do find that the mean \textsf{CROWDSAP} value of our NOV targets (0.35) is 25\% smaller than that of the photometric variables (0.46). One of the NOV systems, TIC 359413177 (J265.3145+29.5881), was identified as an eclipsing HW Vir binary in the EREBOS survey with an orbital period of 0.36103 d \citep{sch19}. Our \tess\ data show no significant photometric oscillations larger than 1 ppt, well below the 10\% variations observed by \citet{sch19} when phase--folding ATLAS photometry. Follow--up observations in future \tess\ sectors might shed further light on these NOV systems. 

\section{Summary \& Conclusions} 
\label{sec:summary}

Using photometric data from \gaia\ DR2, we have identified several hundred candidate variable hot subdwarf stars in the \citet{gei19} catalog from their anomalously high \gmag\ flux errors. We observed 187 of these systems using 2--min cadence \tess\ Cycle 2 observations to test the efficacy of this variability identification method and find over 90\% of the candidates observed to be bona fide variables. It is possible that the targets not observed to vary are also variables, but with amplitudes too low or periods too long to have been detected by \tess. Using a combination of folded light curves, discrete Fourier transforms, positions in the \gaia\ CMD, and, in some cases, follow--up Lick spectroscopy, we were able to classify the majority of observed systems.


 Variable hot subdwarf discoveries in our sample include new HW Vir binaries, reflection effect systems, slowly pulsating sdBV$_s$ stars,  ellipsoidal systems, and a few sdB variables of unknown classification. In some cases, our TESS photometry led to corrected classifications for previously--published systems. Four systems that were identified as candidate eclipsing HW Vir binaries in the EREBOS survey \citep{sch19} show different light curve shapes or no significant variations, including TIC 138025887 (reflection), TIC 440051755 (CV), TIC 402956585 (unclear), and TIC 359413177 (NOV). Additionally, TIC 430960919, which was previously thought to be an sdB+dM/WD binary with 0.81--d period, shows a 0.31--d variation indiciative of a reflection effect and sdB+dM/BD binary. Notably, one of the new reflection effect systems we discovered, TIC 122889490, has the shortest orbital period ever found for a non--eclipsing sdB+dM/BD binary (0.070703 d). Detailed light curve analyses of the HW Vir and reflection effect binaries might lead to hot subdwarf mass estimates and feed studies of the common envelope channel for hot subdwarf formation. Follow--up observations of the new ellipsoidal systems will be needed to confirm their sdO/B+WD nature of this binary and determine whether they are SN Ia progenitors.  For most of our systems --- both new and known --- the \tess\ light curves provide precise timing measurements that might reveal phase oscillations due to additional orbiting objects or evolutionary processes. Many of the variable stars found in our survey turned out {\em not} to be hot subdwarfs at all but instead cataclysmic variables. This result is not entirely unexpected, as the color selection method used by \citet{gei19} extended to much redder colors than single sdB stars in order to include composite systems. 

We also explored our ability to classify variable systems without looking at their folded light curves but instead relying only on the relative amplitudes and phases of the fundamental and first two harmonics in the DFTs of their light curves. From this pilot study, we find that HW Vir binaries, eclipsing CVs, and reflection effect binaries occupy distinct regions of a Fourier diagnostic plot with axes separating their second harmonic amplitudes and first harmonic phases (relative to the fundamental). Other variables, such as non--eclipsing CVs, occupy less distinct regions of this diagram. Nonetheless, investigating the relative amplitudes and phases of the harmonics of a variable's light curve, when combined with its position in the \gaia\ CMD, might permit rapid classification of certain types of binaries. 

In summary, \gaia\ \gmag\ flux errors can serve as powerful indicators of stellar variability and permit efficient follow--up photometry to discover and characterize variable star systems. Similar success was found by \citet{gui20}, who applied the method to white dwarf stars and uncovered several new DAVs and other degenerate variables. We show here that empirical photometric uncertainties can maximize the efficiency and science return on the shortest--cadence, targeted \tess\ observations, and look forward to further monitoring and campaigns from this NASA mission.

\section*{Acknowledgements}

BB, IP, WF, and DV acknowledge funding through NASA Award 80NSSC19K1720. KC, IL, and JH acknowledge support from the National Science Foundation through grant AST \#1812874. 
PN acknowledges financial support from the Polish National Science Center under projects (No.\,UMO-2017/26/E/ST9/00703 and UMO-2017/25/B/ST9/02218), and from the Grant Agency of the Czech Republic (GACR 18-20083S). 
TK acknowledge support from the National Science Foundation through grant AST \#2107982. JJH acknowledges financial support through NASA Award 80NSSC20K0592. BB would personally like to recognize and thank the folks at Camino Bakery in Winston-Salem, NC, for providing wonderful spaces at both their 4th St. and Brookstown locations where the majority of this text was written. He would also like to curse the yellow jackets, Coxsackievirus A16, and broken fourth metacarpal that conspired together to delay the completion and submission of this manuscript.

This paper includes data collected by the TESS mission, obtained through the Cycle 2 Guest Investigator Program \#G022141. Funding for the TESS mission is provided by the NASA Explorer Program. This work has made use of data from the European Space Agency (ESA) mission {\it Gaia} (\url{https://www.cosmos.esa.int/gaia}), processed by the {\it Gaia} Data Processing and Analysis Consortium (DPAC, \url{https://www.cosmos.esa.int/web/gaia/dpac/consortium}). Funding for the DPAC has been provided by national institutions, in particular the institutions participating in the {\it Gaia} Multilateral Agreement. This research has used the services of www.Astroserver.org under reference V429SA. This research has made use of the VizieR catalogue access tool, CDS, Strasbourg, France.

\section*{Facilities:}
\gaia\ (DR2), \tess\ (Cycle 2), Shane 3-m telescope (Kast), ADS, CDS.

\section*{Software:}
ASTROPY (Astropy Collaboration et al. 2013, 2018), IRAF (National Optical Astronomy Observatories), LMFIT (Newville et al. 2014), MATPLOTLIB (Hunter 2007), NUMPY (Harris et al. 2020) PANDAS (pandas development team 2020), PHOT2LC (https://github.com/zvanderbosch/ phot2lc), PHOTUTILS (Bradley et al. 2020), PYRIOD (https://github.com/keatonb/Pyriod), CDS’s (Strasbourg, France) SIMBAD and VizieR online pages and tables, and the NASA Astrophysics Data System (ADS) repositories.

\section*{Data Availability}
The datasets were derived from MAST in the public domain
archive.stsci.edu.


\bibliographystyle{apj}


\newpage
\section*{Appendix A: Other HW Vir Binaries}
\label{appendix:HWVir}

\begin{figure*}[h]
\centering
\includegraphics[width=1.0\columnwidth]{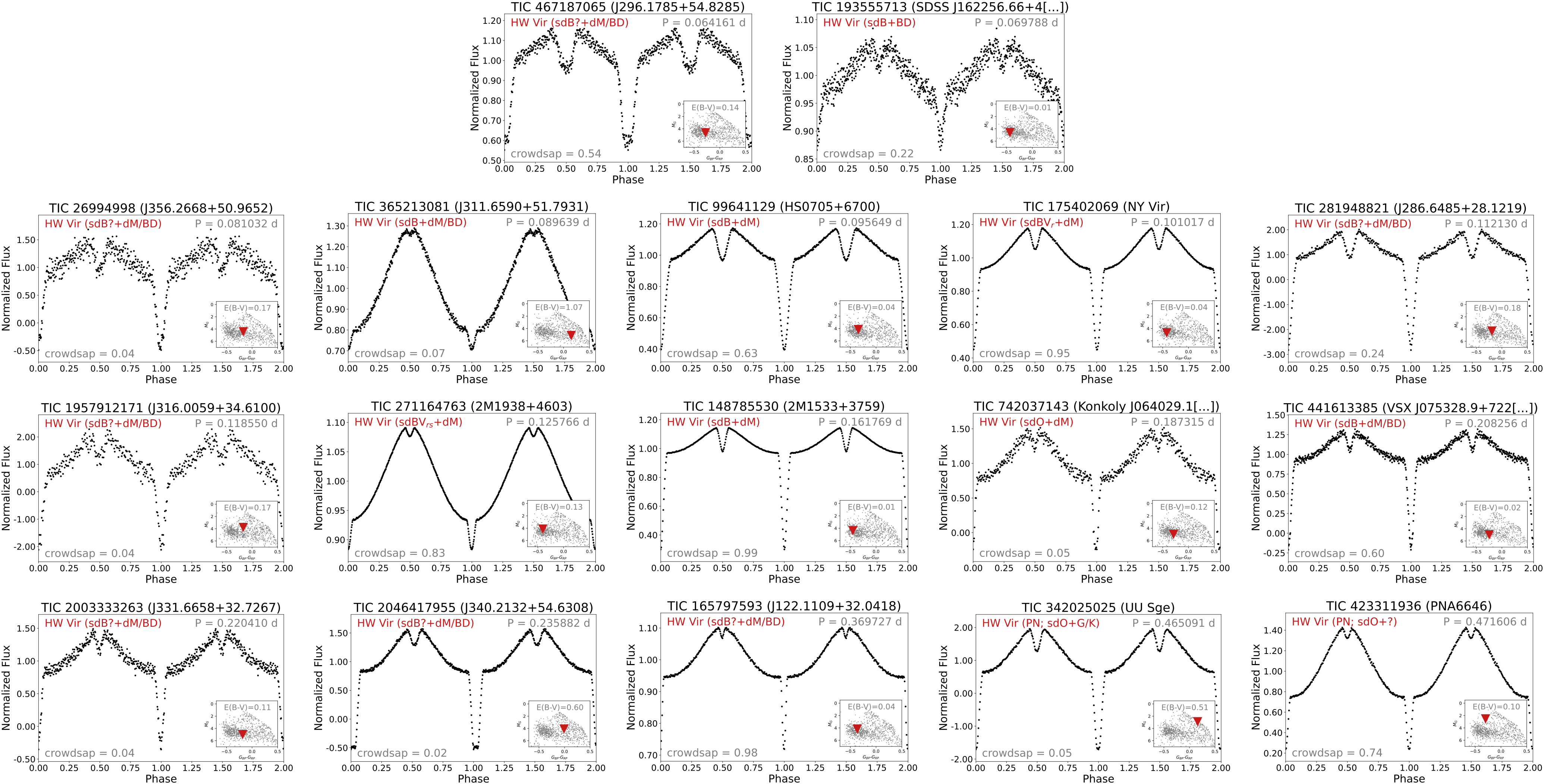}
\caption{\tess\ Cycle 2 phase--folded light curves of other known HW Vir binaries.}
\label{fig:HW_Vir_other}
\end{figure*}

\newpage
\section*{Appendix B: Other New \& Known Reflection Effect Systems}
\label{appendix:reflection}
\begin{figure*}[h]
\centering
\includegraphics[width=1.0\columnwidth]{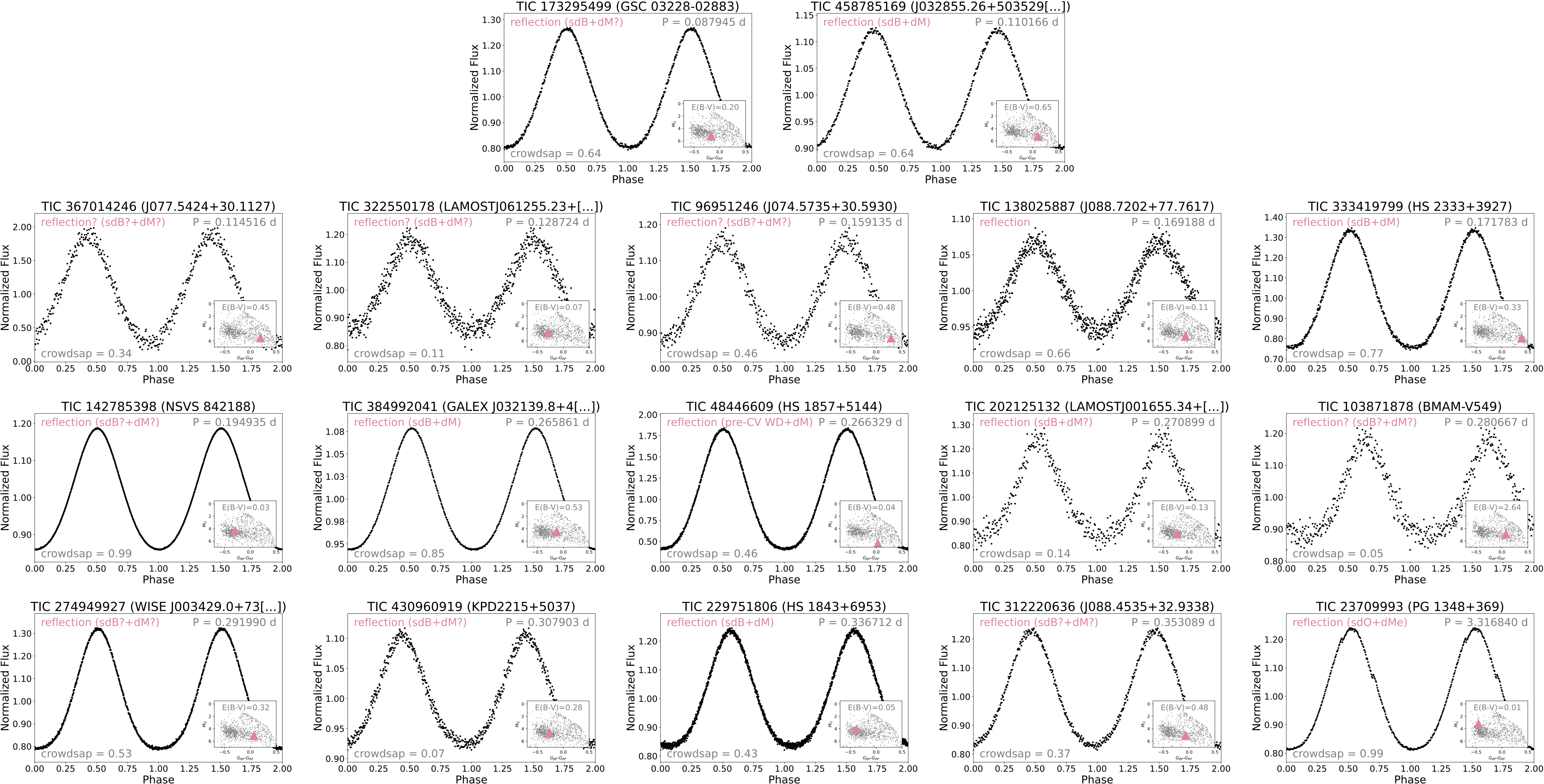}
\caption{\tess\ Cycle 2 phase--folded light curves of other candidate new \& known hot subdwarf reflection effect binaries.}
\label{fig:reflection_other}
\end{figure*}

\newpage
\section*{Appendix C: Eclipsing Cataclysmic Variables}
\label{appendix:eclipsingCVs}

\begin{figure*}[h]
\centering
\includegraphics[width=1.0\columnwidth]{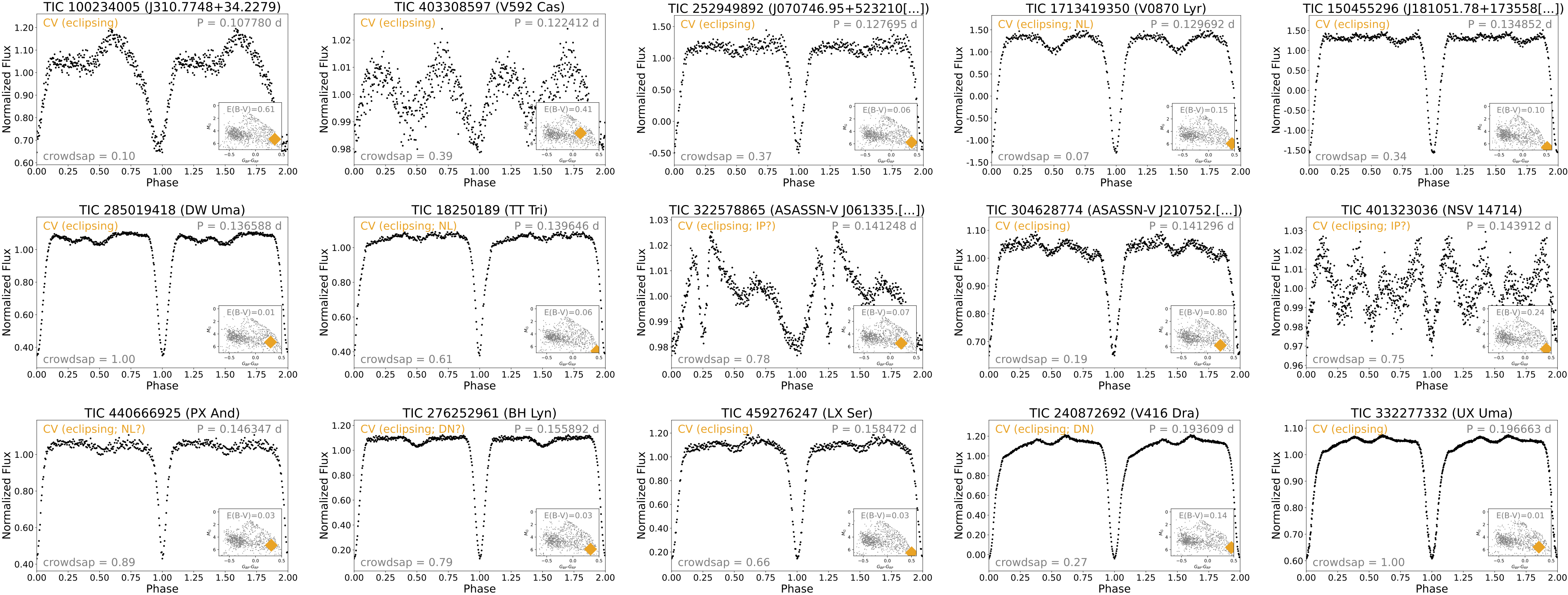}
\caption{\tess\ Cycle 2 phase--folded light curves of other new \& known eclipsing CVs.}
\label{fig:CVs_eclipsing}
\end{figure*}

\newpage
\section*{Appendix D: Other Non-Eclipsing Cataclysmic Variables}
\label{appendix:CVs}

\begin{figure*}[h]
\centering
\includegraphics[width=1.0\columnwidth]{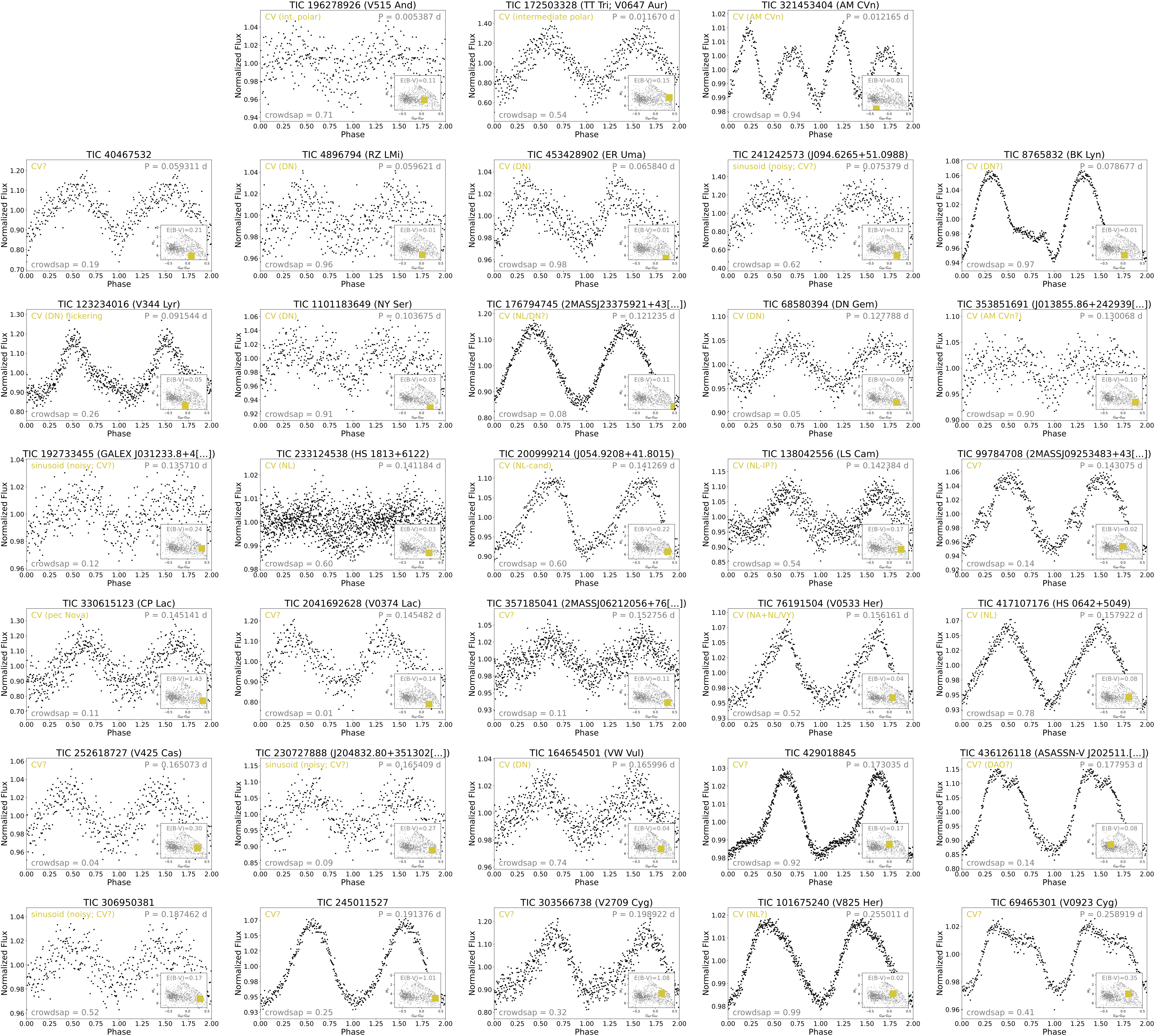}
\caption{\tess\ Cycle 2 phase--folded light curves of other new \& known non--eclipsing CVs.}
\label{fig:CVs_eclipsing}
\end{figure*}

\end{document}